\documentclass[reprint,amsmath,aps,prb,superscriptaddress]{revtex4-2} 

\usepackage{setspace}
\usepackage{lineno} 
\usepackage{widetable}
\usepackage{graphicx} 
\usepackage{braket}
\usepackage{float} 

\usepackage{epstopdf} 

\usepackage{dcolumn} 

\usepackage{hyperref} 
\hypersetup{
	colorlinks   = true, 
	urlcolor     = blue, 
	linkcolor    = blue, 
	citecolor    = blue  
}
\usepackage{placeins}
\usepackage{color} 

\usepackage{amsmath} 

\usepackage{xspace} 
\usepackage{tabularx}
\usepackage{times} 

\usepackage{booktabs} 

\usepackage{lmodern} 

\usepackage{mhchem} 

\newcommand{\add}[1]{\textcolor{black}{#1}}

\usepackage[english]{babel} 

\usepackage{pythonhighlight} 

\newcommand{\DMTTF}{$o$\,-DMTTF\xspace}

\newcommand{\DMTTFX}{($o$\,-DMTTF)$_2X$\xspace}

\newcommand{\Cl}{($o$\,-DMTTF)$_2$Cl\xspace}
\newcommand{\Br}{($o$\,-DMTTF)$_2$Br\xspace}
\newcommand{\I}{($o$\,-DMTTF)$_2$I\xspace}

\newcommand{\etal}{\textit{~et al.}\xspace}

\newcommand{\SI}{
	\setcounter{table}{0}
	\renewcommand{\thetable}{S\arabic{table}}
	\setcounter{figure}{0}
	\renewcommand{\thefigure}{S\arabic{figure}}
	\setcounter{equation}{0}
	\renewcommand{\theequation}{S\arabic{equation}}
	\setcounter{section}{0}
	\renewcommand{\thesection}{S\arabic{section}}
	\setcounter{subsection}{0}
	\renewcommand{\thesubsection}{S\arabic{section}.\arabic{subsection}}
}

\definecolor{bluegray}{rgb}{0.4, 0.6, 0.8}
\newcommand{\bluetitle}{\color{bluegray}}

\newcommand{\affIMNP}{CNRS, Aix-Marseille Universit\'{e}, Université de Toulon, IM2NP, Marseille, France.}
\newcommand{\affBIP}{CNRS, Aix-Marseille Universit\'{e}, BIP , Marseille, France.}
\newcommand{\affLASIR}{CNRS, Universit\'{e} de Lille, LASIRE, Villeneuve d'Ascq, France}
\newcommand{\affISCR}{Universit\'{e} de Rennes, CNRS, ISCR, F-35042 Rennes, France.}
\newcommand{\affISM}{CNRS, Aix-Marseille Universit\'{e},  Centrale Marseille, ISM2,  Marseille, France.}
\newcommand{\affLNCMI}{CNRS, LNCMI (UPR 3228), Laboratoire National des Champs Magnétiques Intenses, EMFL, Grenoble, France.}
\newcommand{\affNEEL}{CNRS, Institut N\'{e}el  (UPR 2940), Grenoble, France.}

\begin{document}
	
\title{Exploring electron spin dynamics in spin chains using defects as a quantum probe.}

\author{L.~Soriano}\thanks{Present address: \affNEEL }\affiliation{\affIMNP}
\author{A.~Manoj-Kumar}\affiliation{\affIMNP}
\author{G.~Gerbaud}\affiliation{\affBIP}
\author{A.~Savoyant}\affiliation{\affIMNP}
\author{R.~Dassonneville}\affiliation{\affIMNP}
\author{H.~Vezin}\affiliation{\affLASIR}
\author{O.~Jeannin}\affiliation{\affISCR}
\author{M.~Orio}\affiliation{\affISM}
\author{M.~Fourmigu\'{e}}\affiliation{\affISCR}
\author{S.~Bertaina}\email{sylvain.bertaina@cnrs.fr}\affiliation{\affIMNP}
\date{\today}

\begin{abstract}
We investigate the quantum dynamics of the electron spin resonance of topological defects (edge
state) in dimerized chains. These objects are discontinuities of the spin chain protected by the
properties of the global system leading to a quantum many-body multiplet protected from the
environment decoherence. Despite recent achievements in the realization of isolated and finite spin
chains, the potential implementation in quantum devices needs the knowledge of the relaxation
and decoherence sources. Our study reveals that electron spin lattice relaxation is governed at lowest
temperatures by phonon-bottlenecked process and at high temperature by the chain dimerization gap. We
show that the inter edge-state effective dipolar field is reduced by the intrachain exchange coupling
leading to a longer coherence time than isolated ions at equivalent concentration. Ultimately, we demonstrate 
that the homogeneous broadening is governed by the intra-chain dipolar field, \add{and we establish design principles for optimizing coherence in future materials}.
\end{abstract}
\maketitle

\section{Introduction}

Topological quantum states in condensed matter systems \add{represent a fascinating class of collective excitations whose properties are governed by global symmetries rather than local perturbations}. A prime example is the S=1/2 topological defect, or "edge state", that emerges in gapped one-dimensional (1D) spin chains \cite{Alloul2009}. While their existence has been established for decades \cite{eggert1992,eggert1994,eggert1995,martins1997}, a fundamental question remains unanswered: what are the intrinsic limits of their quantum coherence, and what physical mechanisms govern their interaction with the environment? \add{Understanding these dynamics is essential both for fundamental physics---to reveal how many-body correlations influence decoherence---and for assessing the potential of such states in future quantum applications.} This paper provides the first comprehensive \add{experimental and theoretical} study of the relaxation and decoherence dynamics of these topological edge states.


Similar effects are predicted in systems undergoing a Spin-Peierls transition systems which have a
nonmagnetic ground state because of a dimerization of the spin and/or lattice
\cite{nishino2000a,sorensen1998,hansen1999}. Furthermore, because the influence of the spins located
at the regular sites of the chains is greatly reduced, a thorough study of the defect-induced
magnetic properties is possible with such a non-magnetic singlet ground state. \add{From a theoretical perspective,} defects in spin chains have attracted considerable interest in the context of quantum information \add{science}. Since Bose's pioneering proposal in 2003 \cite{bose2003}, numerous \add{theoretical} studies have explored the possibility that defects in spin chains could serve as qubits connected by \add{spin-mediated interactions, potentially enabling quantum state transfer protocols} \cite{vieira2020,camposvenuti2006,camposvenuti2007}. \add{While these proposals remain largely theoretical, recent experiments on artificial atomic chains \cite{toskovic2016} and quantum dots \cite{kiczynski2022} have begun to explore related physics. A critical open question for any such application is: what limits the coherence time of embedded defects in real materials? Addressing this question requires a detailed understanding of the microscopic decoherence mechanisms, which has been lacking until now.}

Magnetic
properties of defects in cuprate spin chain like  \ce{Sr2CuO3} \cite{karmakar2015} and \ce{CuGeO3}
\cite{smirnov1998a} have been reported but no coherence was observed so far. Recently, a well-known
class of systems, the organic spin chains with spin-Peierls ground state, has shown edge states with
relatively long coherence times. Initially observed in \ce{(TMTTF)2PF6}\cite{bertaina2014},
the absence of spin echo has prevented the study of decoherence mechanisms. Among the large family of organic spin chains,
one of them is of particular interest: \textit{ortho}-dimethyltetrathiafulvalene or $o$-DMTTF. The serie of
\DMTTFX has been reported with centro-symmetric counter ions (X = Br,Cl and I) \cite{foury-leylekian2011,zeisner2019,soriano2022}.
While these materials show some conductivity at room temperature, they become insulators below about 150~K and undergo a spin-Peierls transition around $T_\text{SP} \simeq 50$ K. Their magnetic properties can be described by the spin-Peierls hamiltonian:

\begin{equation}
	\label{eq:HAFM-Hamiltonian}
	\mathcal{H} = \sum_{i}J(1+\delta)\boldsymbol{S}_{2i-1}\cdot \boldsymbol{S}_{2i} + J(1-\delta)\boldsymbol{S}_{2i}\cdot\boldsymbol{S}_{2i+1} \ \ \ .
\end{equation}
where $J$ is the Heisenberg isotropic coupling, $\delta$ the dimerization parameter and $\boldsymbol{S}_{i}$ a spin $S=1/2$ at the site $i$.
 At high temperature (\textit{i.e.} $T>50~$~K), $\delta=0$ corresponds to the uniform spin chain while at low temperature (\textit{i.e.}
below the spin Peierls transition temperature, $T_\text{SP}$) $\delta \neq 0$ corresponds to the dimerized spin chain with a non-magnetic ground
state. $J$ as well as $\delta$ has been reported using continuous wave EPR  \cite{soriano2022} : $J^{X=Br}=600$~K, $J^{X=Cl}=600$~K,  $J^{X=I}=670$~K , $\delta^{X=Br}=0.08$, $\delta^{X=Cl}=0.083$, and  $\delta^{X=I}=0.093$. Their highly one-dimensional character and the presence of a large spin gap in the bulk provide an ESR-silent background, allowing for an exceptionally clean probe of the in-gap edge states. Our study is performed below 20~K, so we consider that the spin-Peierls phase is fully stabilized. The presence of edge state is du to presence of imperfection in the crystal. In organic materials, this is mainly caused by electrochemical growth, which easily induces local breaks or stacking defects that cannot be corrected by annealing as in inorganic materials. Many
numerical studies \cite{eggert1995,martins1997,sorensen1998,hansen1999,nishino2000a} have shown how
these edge states can influence the overall behavior of the spin chain. In a finite chain, translational
symmetry breaking will polarize the spins in the vicinity of the defect. Using computational
methods such as Density Matrix Renormalization Group (DMRG), we can show that this spin polarization creates a spin cluster whose size
(\textit{\textit{i.e.}} correlation length) depends on the dimerization parameter $\delta$ (see Figure
\ref{fig:SI_localMag_DMRG2}).

\begin{figure}[h!]
	\includegraphics[scale=1]{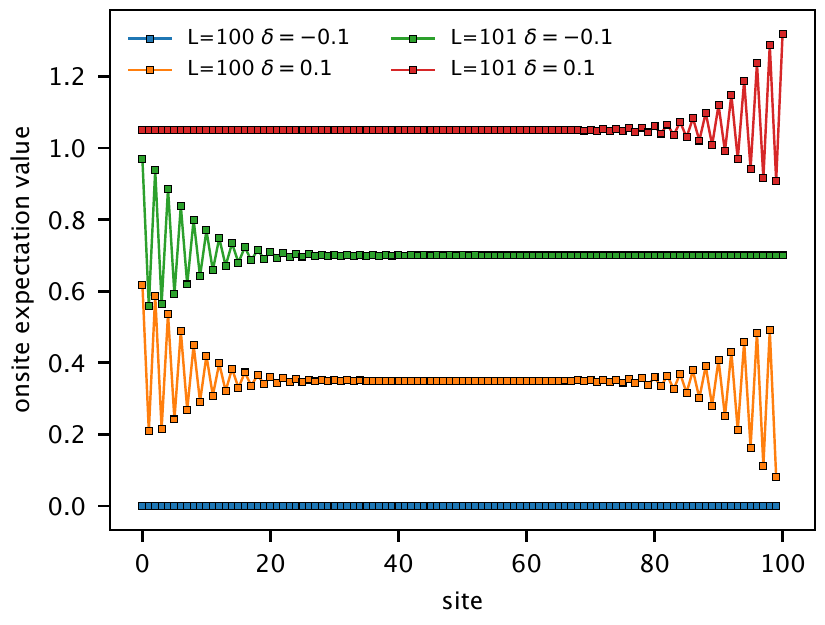}
	\caption{Local magnetization profiles obtained via DMRG
	simulations for open spin chains of length L=100 and L=101 spins. The simulations are performed using the
	spin-Peierls Hamiltonian (eq. \eqref{eq:HAFM-Hamiltonian}) with a dimerization parameter
	$\delta=\pm0.08$. The different colored lines represent the expectation value of the Sz operator at
	each lattice site. The results illustrate how the open boundary conditions lead to a significant
	polarization of the spins at the chain ends, forming a cluster.}\label{fig:localMag_DMRG}
\end{figure}

 For $\delta=0.08$ which is the estimated value for \DMTTFX at low temperature \cite{soriano2022} the
local magnetization of an open spin chain with odd or even number of spins is given in Figure
\ref{fig:localMag_DMRG}. The end of the chain has the effect of polarising many spins forming a
cluster of connected spins to the rest of the chain. This object has a ground state S=1/2
\cite{soriano2020} and is denoted in the literature as of pinned soliton
\cite{uchinokura2002,smirnov1998a,bertaina2014}, end chain \cite{nishino2000a,ikeuchi2017} or edge
state \cite{miyashita1993} and in the following it will be referred as quantum spin chain edge
state (QSC-ES). Due to the topological protection in this many body system, intense effort have
been devoted to create nanoscale individual QSC-ES \cite{fu2025,su2025,zhao2024} but the nature of
decoherence and relaxation mechanism remain unclear.

In this article, we present a comprehensive study of relaxation and decoherence of QSC-ES of a
quasi-1D spin chain embedded in 3D crystals. We have chosen the \DMTTFX family because the bulk magnetic
properties are well known \cite{soriano2020,soriano2022} with a large exchange $J\sim
600$~K and a dimerization parameter $\delta \sim 0.08$ which is neither too small as in \ce{(TMTTF)2X}
family ($\delta\sim 0.03$) \cite{pouget2017} where very narrow homogeneous lines prevented echo observation in previous studies, nor too large  like \ce{BEDT-TTF} \cite{pouget2018} which would lead to overly localized, trivial defects. In
the first part, we show that the spin lattice relaxation of QSD-EC is related the properties of the
bulk spin chain. Then, the decoherence processes of QSD-EC are precisely studied. We demonstrate
that the spin diffusion processes such as instantaneous and spectral diffusions are renormalized by
the many body property of the topologicaly protected QSD-EC.

\section{Experimental Details} 
\subsection{Crystals description.}

\begin{figure}[ht!]
	\includegraphics[width=\linewidth]{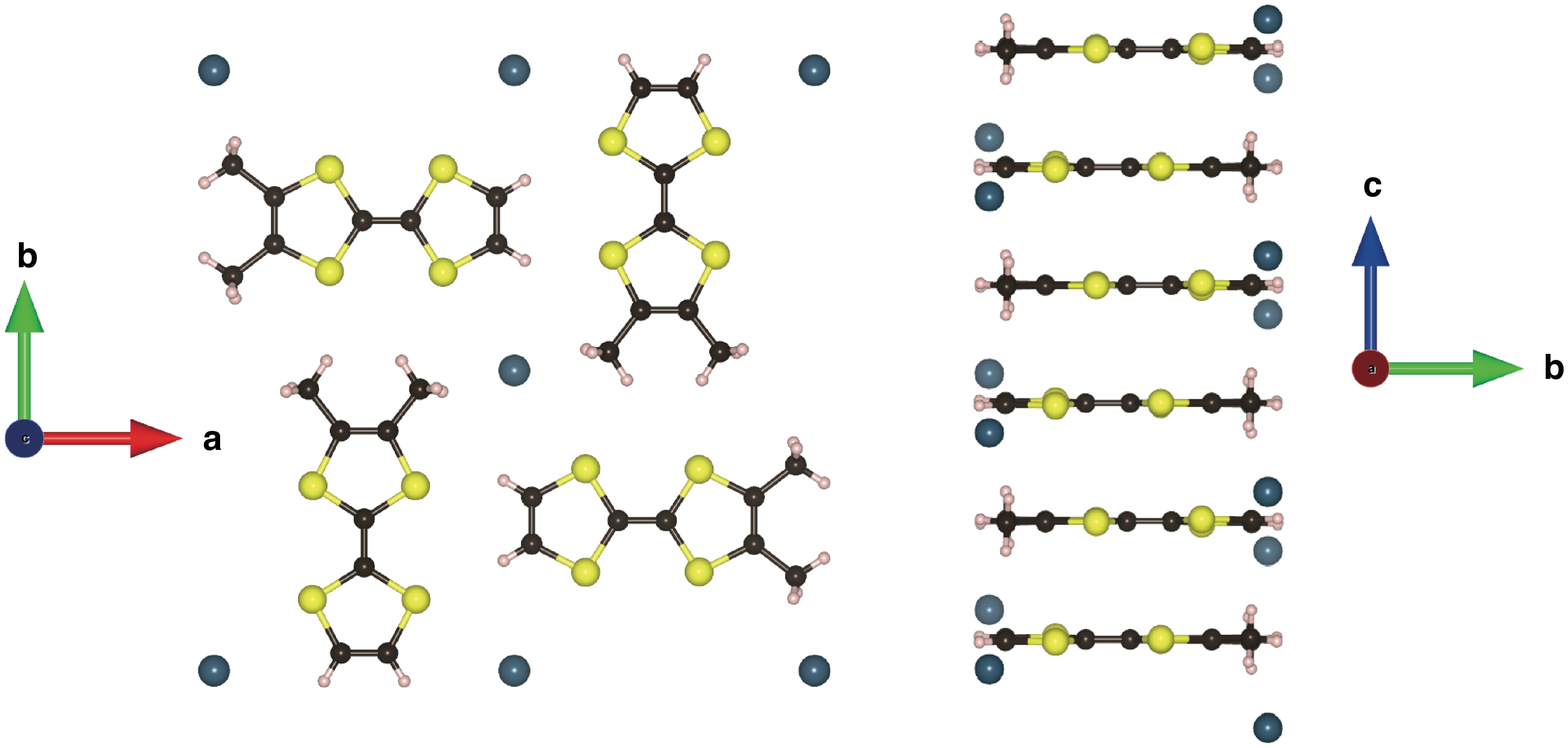}
	\caption{Crystal structures of  \Br, \I and\Cl . The left panels show the unit cell
	viewed along the c axis, while the right panels show the packing along the a axis. The
	crystallographic axes are indicated with colored arrows (red, green, and blue for a, b, and c,
	respectively). The \DMTTF molecules form stacks along the c-axis, with strong intra-stack
	interactions leading to 1D magnetic behavior. The packing and interactions are slightly different
	due to the different counter-anion.}
	\label{fig:Xtalo}
\end{figure}

Single crystals of \Br, \Cl, and \I were grown via electrocrystallization following the standard
procedure described in Ref.\cite{fourmigue2008}. The structural, electronic and magnetic
characterizations were reported in Refs.~\cite{fourmigue2008,foury-leylekian2011,soriano2023}. Their
crystallographic structures are shown in Figure~\ref{fig:Xtalo}. They all have tetragonal unit
cell (space group $I\overline{4}2d$ (no. 122) \cite{fourmigue2008}) with cell parameters $a = b =
16.93 \, \text{\AA}, \, 17.09 \, \text{\AA}, \, 17.40 \, \text{\AA}$ and $c = 7.040 \, \text{\AA},
\, 7.058 \, \text{\AA}, \, 7.098 \, \text{\AA}$, respectively. The cell contains four pairs of molecules of \DMTTF which
stack along the axis perpendicular to the plane of the molecules ($c$ axis). Each pair of \DMTTF molecule is partially oxidized and shares one
hole and so, one spin.  Each stack is rotated by 90$^\circ$ with respect of the neighboring stacks.
As a consequence, the interaction between spins along the stacking axis is very strong
\cite{fourmigue2008,foury-leylekian2011,soriano2022,soriano2023} but the interaction between spin
chains is very weak. The spin-Peierls transition has been reported in all three compounds with a dimerization 
temperature $T_\text{SP}$ around 50~K. Within the temperature range examined in our study ($<$20~K) 
the dimerized phase is fully stabilized.  

These crystals have a needle-like shape with the chain axis along the long length of the needle. The
largest crystals were used in the low-temperature electron spin resonance (ESR) study, where defect signals were observed but
remained weak due to the low concentration of strongly correlated defects. The typical dimensions
were 0.2 $\times$ 0.2 $\times$ 3 mm$^3$, the longest being the chain axis. To avoid the effects of
temperature cycle history, a fresh sample was used for each series of measurements. The samples were
glued to a suprasil quartz rod using a small amount of Apiezon grease on one side to minimize stress
during temperature sweeps. The samples came from the same batches that the ones used for cw-ESR concentration obtained in \cite{soriano2022} and this in order to avoid concentration difference between samples. Moreover, to be sure about the concentration we have re-performed a cw-ESR intensity measurements right after the ID study. 

\subsection{ESR spectrometer setup}
Coherent spin dynamics were investigated using a pulsed X-band (9.4 GHz), Q-band (34GHz) and W-band
(94GHz) ESR spectrometer (Bruker E580), equipped with an Oxford CF935 He-flow cryostat and a
ColdEdge Stinger cryogen-free coldhead, allowing to reach a base temperature of 5.5~K at the sample
position. The experiments were conducted using microwave cavities optimized for pulsed operation.
We performed inversion
recovery experiments (\(\pi - t_1 - \pi/2 - \tau - \pi - \tau - \text{echo}\)) for $T_1$
measurements, and used 2-pulse Hahn echo  (\(\pi/2 - \tau - \pi - \tau - \text{echo}\)) and
3-pulse  (\(\pi/2 - \tau - \pi/2 - T_W - \pi/2 -\tau - \text{echo}\)) stimulated echo sequences for
$T_2$ measurements. The repetition rate was set in order to let the spin system fully relax.

\section{Spin Lattice Relaxation}\label{sec:SLR}

The spin-lattice relaxation time, $T_1$, represents the characteristic duration required for a spin
to return to its equilibrium state after random flips along the spin quantization axis. Longitudinal relaxation $T_1$
imposes an intrinsic limit on the transverse relaxation $T_2$, such that $T_2 \leq 2T_1$. Therefore, understanding the mechanisms
 of spin-lattice relaxation is crucial for understanding spin decoherence. In magnetically diluted systems,
  such as rare earth ions, transition metal
ions or nitrogen vacancy (NV) centers, which represent the majority of the studies on electron spin qubit, the spin-flip
induced by the spin-phonon interaction is well understood and can be divided in three processes
(Fig.~\ref{fig:SI_diagram}): The direct absorption or emission of a phonon resonating between the
spin states $\left| a\right\rangle $ and $\left| b \right\rangle $. This process
dominates at low temperature, when the density of phonon is mostly at the energy of the microwave
$h\nu = \delta$. When the temperature increases, thermally assisted relaxation mechanism using two
phonons is induced: The non resonant Raman process occurs by virtual absorption of a phonon of energy
$\delta_1$ and is followed by emission of a phonon of energy $\delta_2$ and finally, the resonant
Orbach process occurs through a third state $\left| c \right\rangle $. This last mechanism exists only if the Debye temperature $\theta_D$ is larger than the energy of the third state $\left| c \right\rangle $ . For a Kramers ion, the spin-lattice relaxation time as a function of temperature is given by \cite{abragam2013,shrivastava1983}:
 
\begin{equation}\label{eq:SLR-process}
	\begin{split}
		\frac{1}{T_1} &= \alpha_D \cdot T + \alpha_{R5} \cdot T^5+\alpha_{R7} \cdot T^7+ \\
		&\quad \alpha_{R9} \cdot T^9 + \alpha_O \cdot\Delta_0^3 \exp{(-\Delta_0 / k_B T)}
	\end{split}
\end{equation}          
where $\alpha_D$ is the direct absorption-emission process coefficient and is dependent to the
external magnetic field as $H_0^4$, $\alpha_{Ri}$ are the two-phonon Raman process coefficients such as
$\alpha_{R5}$ describes mostly molecular systems, $\alpha_{R7}$ is proportional to $H_0^2$,
$\alpha_{R9}$ is the principal coefficient expected in Kramers ions, $\alpha_{O}$ is the resonant
Orbach coefficient\cite{orbachr.1961} with $\Delta_0$ the gap with the first excited spin state.

\begin{figure*}[t!]
	\centering
	\includegraphics[width=\textwidth,clip]{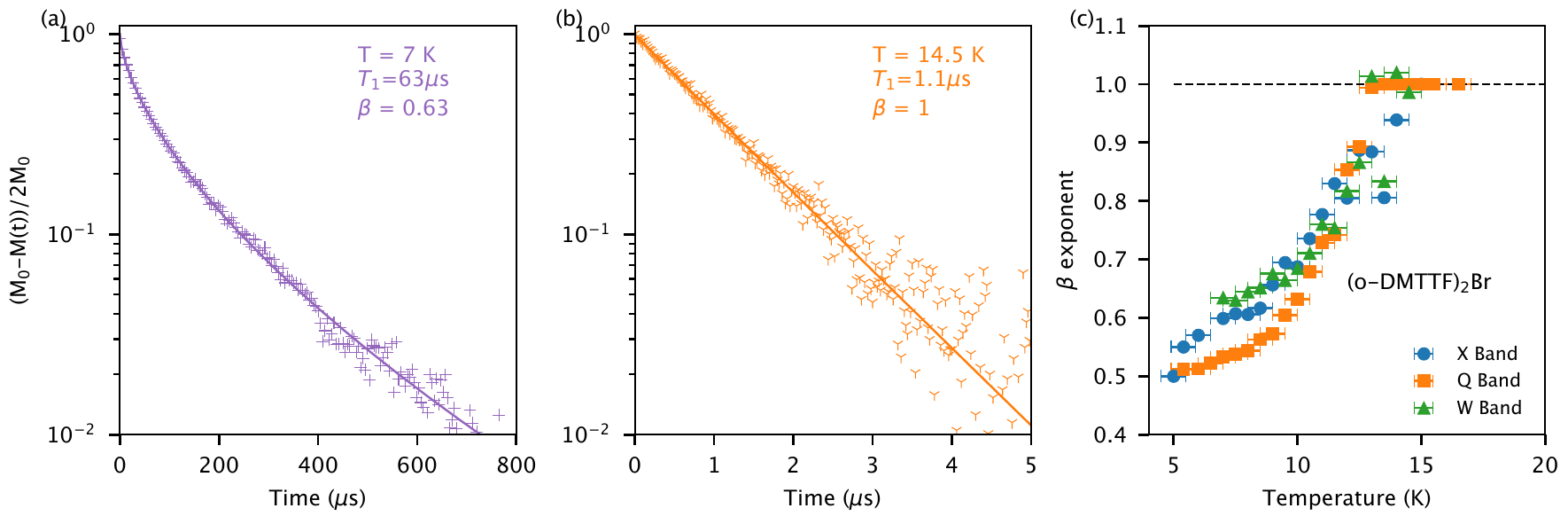}
	\caption{(a)-(b) Longitudinal magnetization as function of time after the inversion pulse. The
	semi-log scale is used to highlight the deviation from the monoexponential decay. The solid
	line is the best fit using eq.~\eqref{eq:fit_relaxation}. (a) At $T=7$~K, the stretched exponential
	parameter is $\beta=0.65$ and (b) at $T=14.5$~K $\beta=1$. (c) Temperature dependence of $\beta$ for \Br and
	for the three microwave bands used.}
	\label{fig:InvRecovery_beta}
\end{figure*}

In this section we study $T_1$ as function of the temperature between T=5~K and 20~K and as function of
the magnetic field using three microwave frequencies  9.7~GHz (0.35~T), 34~GHz (1.21~T) and 94~GHz (3.35~T)
using the inversion recovery pulse sequence: A $\pi$ pulse inverses the population of spins, and after
waiting for a time $t_1$ the recovered magnetization $\left<S_z\right>$ is probed using a Hahn echo ($\pi -
t_1 - \pi/2-\tau-\pi-\tau-\text{echo}$). Examples of magnetization recovery at $T=7$~K and 14.5~K are
shown in Figure~\ref{fig:InvRecovery_beta}. While at high temperature, the recovery is clearly
monoexponential, it deviates at low temperature and a stretched exponential
is needed to fit the experimental values (eq.\eqref{eq:fit_relaxation}).

\begin{equation}\label{eq:fit_relaxation}
	\frac{M_0-M_z(t)}{2} = \exp{\left(-t/T_1^\text{str}\right)^{\beta}}
\end{equation}

The variation of the stretched parameter $\beta$ as function of temperature for the three
frequencies used is shown for \Br in Figure~\ref{fig:InvRecovery_beta} (c). At high temperature
($T>13$~K), $\beta=1$ indicates a uniform distribution of $T_1$ while below ($T<13$~K), $\beta$
progressively decreases with a minimum of 0.5 at $T=5$~K. No significant effect of the microwave
frequency, and consequently, of the magnetic field is observed.

The stretched exponential behavior of longitudinal magnetization has been observed in a wide variety
of systems and research areas \cite{berberan-santos2005,stein2019,phillips1996}, and usually fits 
relaxation processes in disordered and quenched electronics. However the microscopic physical meaning
of $T_1^\text{str}$ when $\beta < 1$ is debatable. Johnston \cite{johnston2006} has shown that the
stretched exponential is the Laplace transform of the probability density $P(1/T_1,\beta)$ of
$1/T_1$. Knowing $\beta$ it is then possible to reconstruct $P(1/T_1,\beta)$ (see Figure~\ref{fig:SI_relacdistrib} for some examples). 
It is worth to notice that the maximum of  $P(1/T_1,\beta)$ is not $T_1^\text{str}$ but
Johnston has shown that in the case of $0.5<\beta<1$, $1/T_1^\text{str}$ can be described as the median of
$P(1/T_1,\beta)$. Thus, while the temperature dependence of $T_1^\text{str}$ has a physical significance,
extracting and interpreting quantitative prefactors could be hazardous.

\begin{figure}[tbh!]
	\includegraphics[scale=1]{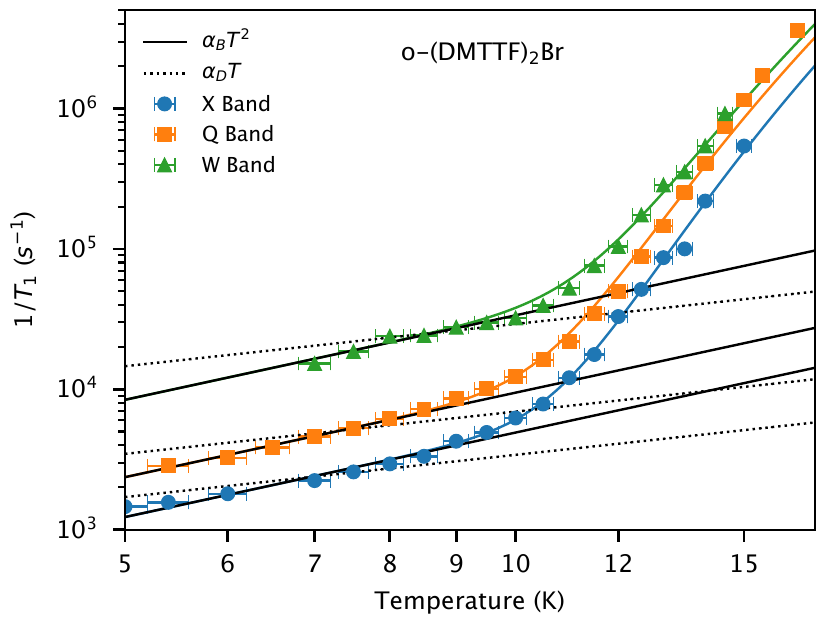}
	\caption{Temperature dependence of the inverse stretched relaxation time ($1/T_1^\text{str}$) for
	\Br, obtained via inversion recovery experiments at three microwave frequencies: X-band (9.7 GHz),
	Q-band (34 GHz), and W-band (94 GHz). The data are plotted on a log-log scale. Dashed lines
	represent fits to a linear temperature dependence ($\alpha_D T $), reflecting the direct phonon
	process, which provides a poor fit. Black solid lines show fits to a square of temperature 
	dependence ($\alpha_B T ^2$), indicative of a possible phonon bottleneck effect.}
	\label{fig:SLRvsT_LT}
\end{figure}

\begin{figure}[t!]
	\includegraphics[scale=1]{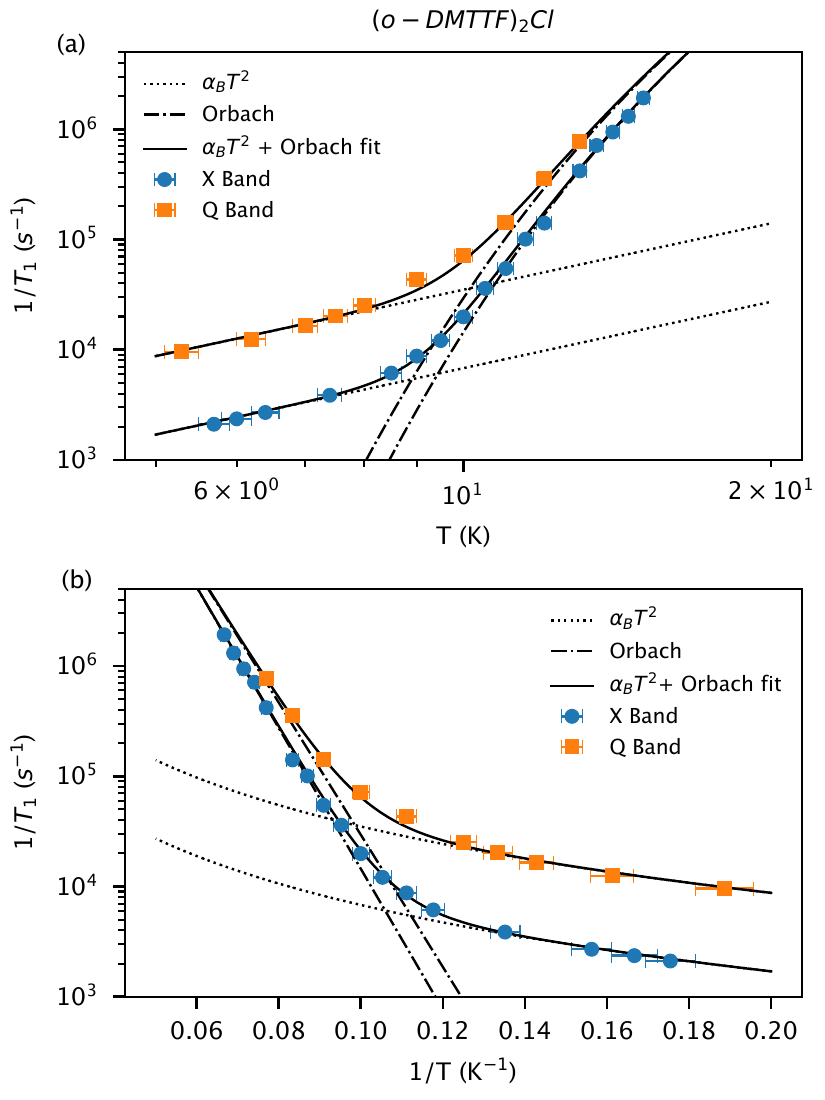}
	\caption{Temperature dependence of the inverse stretched relaxation time ($1/T_1^\text{str}$) for
	\Cl, plotted against temperature (a) or the inverse of temperature (b) in a log-log scale, and
	measured at two microwave frequencies: X-band (9.7 GHz) and Q-band (34 GHz). The data are obtained
	via inversion recovery experiments at variable temperatures between 6 K and 20 K. The dashed lines
	represent the high temperature Orbach process and the dot line the phonon-bottleneck direct
	process. The plain black line is the sum of the two contribution}\label{fig:SI_T1_Cl}
\end{figure}

\begin{figure}[t!]
	\includegraphics[scale=1]{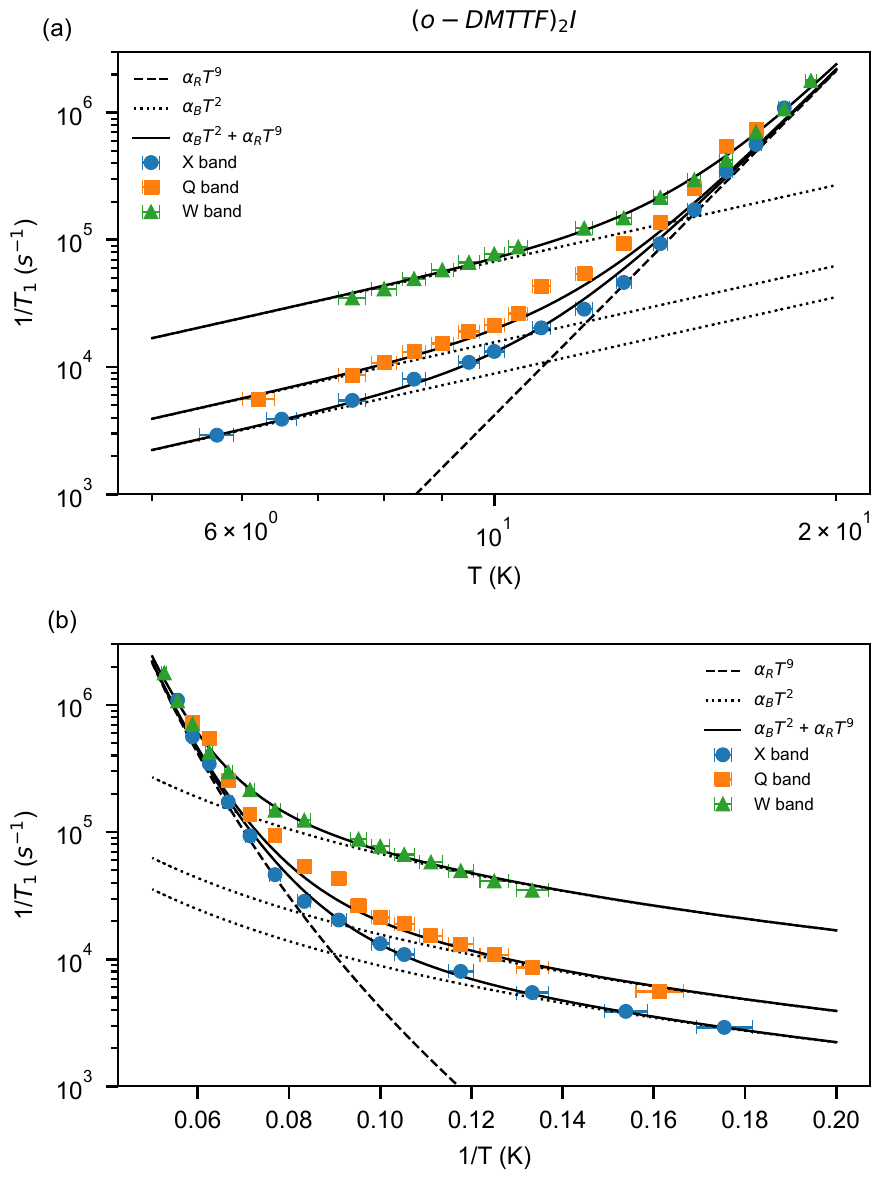}
	\caption{Temperature dependence of the inverse stretched relaxation time ($1/T_1^\text{str}$) for
	\I, plotted against temperature (a) or the inverse of temperature (b) in a log-log scale, and
	measured at three microwave frequencies: X-band (9.7 GHz), Q-band (34 GHz), and W-band (94 GHz).
	The data are obtained via inversion recovery experiments at variable temperatures between 6 K and
	20 K. The dashed lines represent the high temperature Raman process and the dot line the
	phonon-bottleneck direct process. The plain black line is the sum of the two
	contribution}\label{fig:SI_T1_I}
\end{figure}

Now, we examine the temperature dependence of $1/T_1^\text{str}$. Using the case of \Br, Figure~\ref{fig:SLRvsT_LT}
illustrates the temperature dependence of $1/T_1^\text{str}$ in log-log
scale  for the three microwave frequencies available. The dashed line is the best fit for the low
temperature part ($T \lesssim 11$~K) using the direct one-phonon process ($\alpha_D.T $) which is
assumed to be dominant in this range of temperature. Clearly the fit is poor and it is evident that
the relaxation rates $T_1^\text{str}$ do not exhibit linear temperature dependence. This
observation holds true for \Cl (Fig.~\ref{fig:SI_T1_Cl}), and \I (Fig.~\ref{fig:SI_T1_I}) 
 as well. A much better fit was found using a square temperature dependence of
$1/T_1^\text{str}$ using $\alpha_B.T ^2$ (black lines in Fig.~\ref{fig:SLRvsT_LT}). The magnetic
field dependence of the spin-lattice relaxation does not follow the usual behaviour of the direct
regime. In fact, in a Kramers doublet,such as in our case, we would expect a dependence in $H_0^4$
and therefore a variation of 4 orders of magnitude between the X-band and W-band measurements, which
is clearly not the case. Here, the low-temperature behaviour is rather linear with the magnetic
field.

The linear field dependence and the quadratic temperature dependence of $1/T_1^\text{str}$ suggest
that the direct process is "bottlenecked". Indeed, in the direct-phonon process, it is assumed that
the phonon emitted during the relaxation is rapidly absorbed by the thermal bath. However, if the
phonon remains long enough in the crystal, it might be re-absorbed by the spin system, reducing the relaxation
efficiency. The phonon-bottleneck was predicted by van Vleck \cite{vanvleck1941}
and observed in various systems. The relaxation rate in a phonon-bottlenecked direct process is
approximated by \cite{abragam2013,scott1962}:

  \begin{equation}
  	\label{eq:phonon_bottleneck_rate}
  	\frac{1}{T_1^{\mathrm{phb}}} = \frac{3 \omega^{2} \Delta \omega}{2 \pi^{2} v^{3} n \tau_{\mathrm{ph}}} \operatorname{coth}^{2}\left(\frac{\hbar \omega}{2 k_{\mathrm{B}} T}\right)
  \end{equation}
  
where $v$ the velocity of sound in the crystal, $\Delta \omega$ is the resonance linewidth, $n$ is
the concentration of spins, and $\tau_{\mathrm{ph}}$ is the lattice–bath relaxation time. Assuming
$k_BT \gg \hbar\omega$, and the inhomogenous linewidth is field dependent we can simplify eq.\ref{eq:phonon_bottleneck_rate} by

  \begin{equation}
	\label{eq:phonon_bottleneck_rate2}
	\frac{1}{T_1^{\mathrm{phb}}} = \alpha'_B H_0 T^2
\end{equation}
It has been observed that the phonon-bottleneck process of the spin lattice relaxation is
accompanied by a non-exponential decay of the magnetization \cite{budoyo2018} (see eq.\eqref{eq:fit_relaxation})
 which is a behavior we also observed at low temperature. We can conclude
that the temperature dependence of the stretched parameter $\beta$ is the signature of the
phonon-bottleneck process that takes place below about 13~K (see Fig.~\ref{fig:InvRecovery_beta}(c)).
 
 \begin{table*}[htb!] \caption{Fitted parameters of spin lattice relaxation.$\alpha_B=\alpha_B'H_0$ (see \eqref{eq:phonon_bottleneck_rate2}) is the direct-bottleneck $\alpha_{R9}$ the Raman and $\alpha_{O}$ the Orbach coefficients. The fit error estimation is included in the last digit.   \label{tab:T1}}
	
	\begin{widetable}{\textwidth}{lcccccc} \hline
		\noalign{\smallskip}\hline
		\noalign{\smallskip}
		& Band &  $\alpha_B$  & $\alpha_{R9}$ & $\alpha_O$ & $\Delta_0$&$\Delta_\text{SP}$ \cite{soriano2022,soriano2023} \\
		&  &  $(\text{s}^{-1}\text{K}^{-2}$) & $\times 10^{-6}(\text{s}^{-1}\text{K}^{-9}$)& $\times 10^{4}(\text{s}^{-1}\text{K}^{-3})$& (K)& (K)\\
		\noalign{\smallskip}\hline\noalign{\smallskip} 
		\text{\Cl}& X & 68 & - & 1.1 & 147& 178 \\
		& Q & 350 & - & 1.1 & 138& 178 \\
		
		\text{\Br}	& X & 49.1 & - & 1.4 & 181 &181\\
		& Q & 94.6 & - & 1.4 & 170&181\\
		& W & 336 & - & 1.4 & 165&181\\
		
		\text{\I} & X & 88.8 & 4.17 & -& -&230\\ 
		& Q & 157 & 4.17 & -& -& 230\\ 
		& W & 676 & 4.17 & -&-& 230\\ 

		\noalign{\smallskip}\hline\hline 
		\end{widetable}\label{Tab:T1} 
\end{table*}

\begin{figure}[h!]
	
	\includegraphics[width=\linewidth]{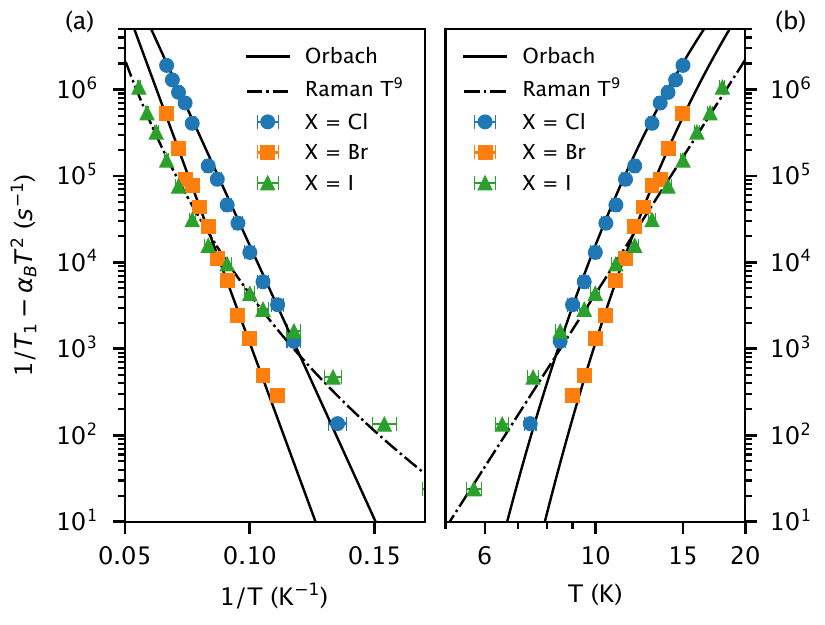}
	\caption{Temperature dependence of the spin lattice relaxation rate corrected by the phonon bottleneck contribution  with respect to (a) $1/T$ or (b) $T$ in  log-scale. Plain and dash-doted lines are the best fit using Orbach and Raman model respectively. }
	\label{fig:SLRvsT_HT}
\end{figure}

Having identified the low-temperature phonon-bottlenecked direct process, we now focus on the behavior at higher
temperature. To study the two-phonon regimes, we have subtracted the low-temperature
behavior from the total relaxation : $1/T_1^{HT}=1/T_1-\alpha_BT^2$. Since the Orbach- and Raman-processes are highly non-linear, we present in Figure~\ref{fig:SLRvsT_HT} the temperature dependence of
the corrected relaxation either as ($i$) $\log(1/T_1^{HT})$ vs $1/T$ or ($ii$) $\log(1/T_1^{HT})$ vs
$\log(T)$. In the first case ($i$) the Orbach (resonant) process is linear with respect of the
inverse of temperature while in the second case ($ii$), the Raman (non-resonant) process is linear in
the log-log scale with the slope corresponding to the exponent (see eq. \eqref{eq:SLR-process}).

The best fit using pure Orbach : $1/T_1=\alpha_{O}\Delta_0^3\exp(-\Delta_0/k_BT)$ and pure Raman :
$1/T_1=\alpha_{R9}T^9$ shows two different groups of systems. \Cl and \Br exhibit an Orbach process
while \I has Raman-$T^9$ process. We think that these two different relaxation behaviors are
directly related to the relative value of the spin-triplet gap. 
In the cases of \Cl and \Br, the Orbach gaps $\Delta_0$ obtained from the best fit of the temperature dependence of $T_1$ (Fig.~\ref{fig:SLRvsT_HT}) is coherent with the value of the spin-triplet gap $\Delta_\text{SP}$. However, in the case of \I , $\Delta_\text{SP}$ is large and no Orbach process is observed suggesting that no phonon has the energy to reach the triplet state (\textit{i.e.} $\theta_D>\Delta_\text{SP}$). Summary of the results for the three compounds and the accessible frequencies is reported in Table \ref{Tab:T1}.

\section{Electronic spin coherence time } 

In this part, we focus on the electronic coherence time
$T_2$ and the mechanisms of decoherence of \I, \Br and \Cl. Several phenomena contribute to the loss of coherence of
electronic spins. Firstly, the spin-lattice relaxation resulting from spin-phonon interactions sets
the physical limit for the coherence time, \textit{i.e.}, $T_2<2T_1$. We have discussed the various
relaxation mechanisms in detail in section \ref{sec:SLR} and identified which ones were involved in
the relaxation of \DMTTFX. This limit is rarely reached because the excited spins are subject to
random dephasing caused by spin-spin interactions. These dephasing processes contribute to reducing
the coherence time $T_2$ and are collectively represented by the term $\Gamma_{\phi}$. The coherence
time is then related to the relaxation time $T_1$ by the relation:

\begin{equation}\label{eq:T2GammaT1} 
	\frac{1}{T_2} = \Gamma_{\phi}+\frac{1}{2T_1} 
\end{equation}

Let us note that the term $\Gamma_{\phi}$ is a sum of different contributions to the random phase
shifts. Some contributions to the phase shifts are intrinsic, such as the flip-flop process, while
others are due to the environment and may depend on the measurement method, such as spectral
diffusion \cite{mims1961} and instantaneous spectral diffusion \cite{klauder1962,raitsimring1985}.
Therefore, the coherence time value depends on the measurement method. These various decoherence
mechanisms have been detailed in \cite{Wolfowicz2021}.

We employ both two-pulse Hahn echo (2PE) and three-pulse stimulated echo (3PE) to clarify the
dynamic interactions that give rise to spin diffusion (SD) in QSC-ES ensembles. In this context, the
2PE involves flipping the spins into the transverse plane with a $\pi/2$ pulse. The spins defocus
for a time $\tau$, and then we refocus them with a pulse corresponding to an angle $\theta_2$
(usually optimum for $\theta_2=\pi$), the total sequence being $\frac{\pi}{2} - \tau - \pi - \tau -
\text{echo}$. The 3PE splits the refocusing $\pi$ pulse into two $\frac{\pi}{2}$ pulses separated
by a time $T_W$: $\frac{\pi}{2} - \tau - \frac{\pi}{2} - T_W - \frac{\pi}{2} - \tau -  \text{echo}$.
The 2PE and 3PE sequences were first used by Mims \etal \cite{mims1961} and since then the underlying theory 
was improved using uncorrelated-sudden-jump \cite{hu1974,bottger2006}:
 
 \begin{equation}\label{eq:I3p}
 	I_E(2\tau+T_W) = V_{\text{ESEEM}} I_0 \exp \left( -2\pi \Gamma_{\text{eff}} \tau \right) \exp \left( -\frac{T_W}{T_1} \right)^\beta
 \end{equation}
 
 \begin{equation}\label{eq:Gamma_eff}
 	\Gamma_{\text{eff}} = \Gamma_0 + \frac{\Gamma_{\text{SD}}}{2} \left\{ R \tau + 1 - \exp \left( -R T_W \right) \right\}
 \end{equation}

where $I_0$ is the echo amplitude at zero delay, and $V_{\text{ESEEM}}$ represents to the electron
spin echo envelope modulation (ESEEM) caused by the magnetic dipole-dipole interaction of the QSC-ED
and the nuclear spins in \DMTTFX \cite{mims1972}. ESEEM from the
host nuclear spins such as $^1$H, $^{13}$C and counter-ions was not observed. This absence of signal will be discussed
after. $T_1$ is the spin-lattice relaxation and we have added $\beta$ to take into account the
phonon-bottleneck effect presented before. Eq. \eqref{eq:I3p} and \eqref{eq:Gamma_eff}  are valid when $\tau\ll T_1$.

The effective decaying rate $\Gamma_\text{eff}$ incorporates the coefficients \(\Gamma_{0}\),
\(\Gamma_\text{SD}\) and \(R\). The coefficient \(\Gamma_{0}\) characterizes decoherence processes
fast enough compared to the timescale of the measurements and also the instantaneous diffusion
induced by the microwave excitation pulses. \(\Gamma_\text{SD}\) and \(R\) describe the spectral
diffusion during the delays \(\tau\) and \(T_{W}\).  Specifically, \(R\) denotes the average
spin-flip rate, while \(\Gamma_\text{SD}\) represents the full width at half maximum (FWHM)
contribution to the dynamic distribution of transition frequencies.

In the case of $T_{W} = 0$ the sequence corresponds to the 2PE.

\begin{align}
	I(2\tau) &= I_0 \exp\left(-2 \pi \Gamma_{0}\tau  -2 \pi\frac{\Gamma_\text{SD}R}{2}\tau^2 \right) \label{eq:2PE_part1} \\
	&= I_0 \exp\left(- \left(\frac{2\tau}{T_2}\right)^n \right) \label{eq:2PE_part2}
\end{align}

In the presence of pure spectral spin diffusion caused by non resonant spin bath (nuclear or
electronic), the decay is expected to be a stretched exponential with $n$ ranging between 2 and
3 \cite{mims1961,klauder1962,chiba1972}. A comprehensive derivation as well as validity questions can be found in Ref. \cite{ledan2022}.

\subsection*{Instantaneous diffusion}

Instantaneous diffusion (ID) is a decoherence mechanism caused by the microwave excitations used to
refocalize the spins that generate a spin echo. During a $\pi$ refocusing pulse, spin populations
are inverted, altering the quantum state m$_s$ of the resonant spins, and the locally perceived
field for the spins is abruptly changed. The dephasing generated by ID is related to the coherence
time by eq.\eqref{eq:ID1}:

\begin{equation}\label{eq:ID1} 
	\frac{1}{T_2} = \pi \Gamma_{0,ID}  +
	\pi \Gamma_{ID}\left<\sin^2{\theta_2/2}\right>_f 
\end{equation} 
where the phase shift term $\Gamma_{ID}$ of eq.\eqref{eq:ID2} is proportional to the spin concentration and
$\Gamma_{0,ID}$ is the decoherence in absence of ID such as homogeneous broadening and spectral
diffusion (SD).

\begin{equation}\label{eq:ID2}
	 \Gamma_{ID} = \frac{\mu_0 g^2 C \mu_B^2 }{9\sqrt{3}\hbar}
\end{equation}

A key characteristic is the angle of rotation of the refocalisation pulse
$\theta_2$\cite{agnello2001,boscaino1992a} through the expression $\left<\sin^2{\theta_2/2}\right>_f$,
which describes the average probability of inverting all spins. It depends on various parameters
such as the microwave field (proportional to the Rabi frequency $\nu_R$) and the profile $f$ of the ESR
line (a lorentzian of half width at half maximum $\Delta H_{1/2}$ in our case):
 
\begin{equation} 
	\left<\sin^2{\theta_2/2}\right>_f = \int_{-\infty}^{+\infty}
	\frac{\nu_R^2}{\nu_R^2+\eta^2}\sin^2{\left(\frac{\theta_2}{2}\sqrt{1+\frac{\eta^2}{\nu_R^2}}\right)} f(\eta+\Delta)d\eta
\end{equation}

When $\left<\sin^2{\theta_2/2}\right>_f$ = 1, all spins involved in the ESR line are excited. 
However, in our case the effect of a partially excited line has to be taken into account. As an
example let's consider \I where $\nu_R=10.5$~MHz and $\Delta H_{1/2}=$ 2.5~G ($\equiv$ 7~MHz). With
our parameters, $\left<\sin^2{\theta_2/2}\right>_f$ = 0.56 for a total refocusing pulse $\theta_2$ =
180°. This situation occurs when the microwave power is not sufficient to excite all spins
constituting the inhomogeneous line.

\begin{figure}  
	\includegraphics[scale=1]{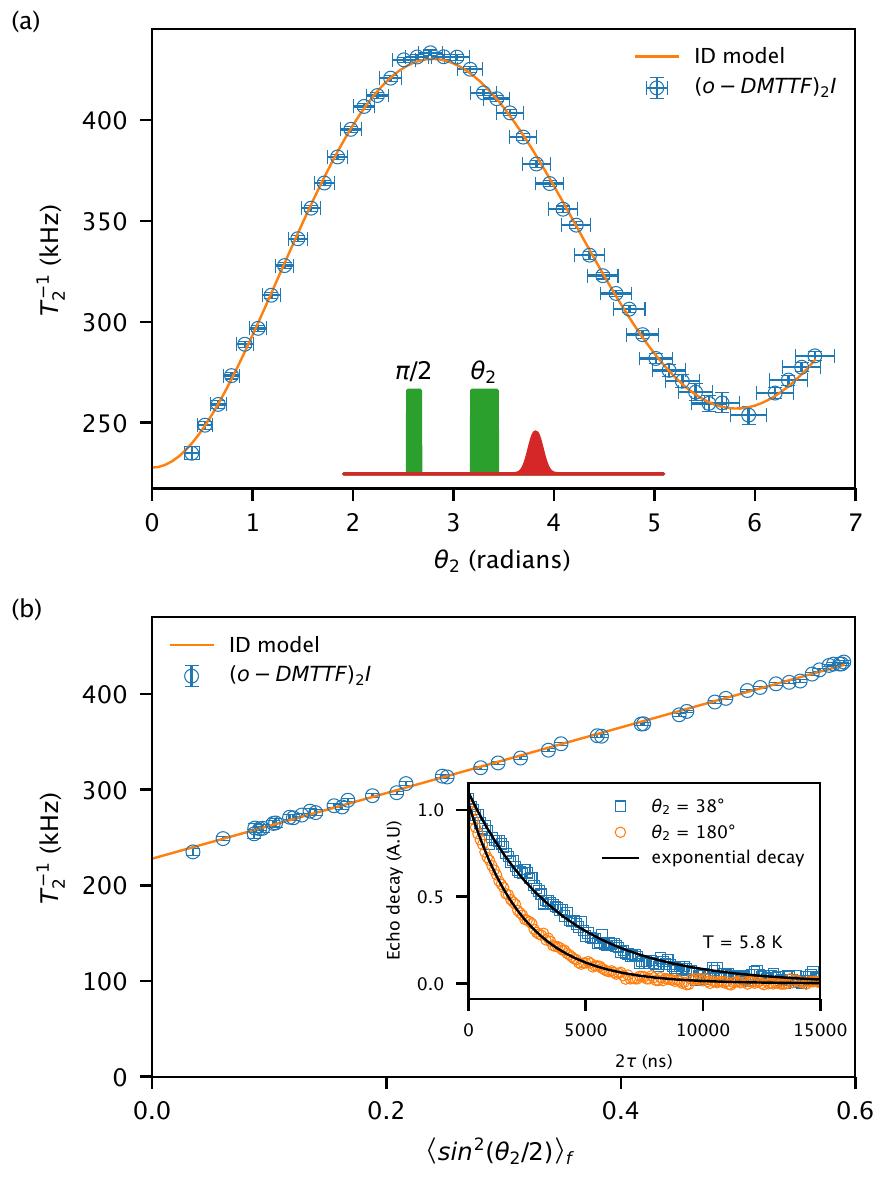}
	 \caption{Effect of instantaneous diffusion on electronic coherence time $T_2$ measured by Hahn echo sequence of
		\I, $T = 5.8$~K, $H_0\parallel c$ and in X band. (a) Plot of $1/T_2$ as a function of refocusing
		angle $\theta_2$ and (b) as a function of $\left<\sin^2{\theta_2/2}\right>_f$ where $\theta_2$ is rotation angle of the second pulse in the sequence $\frac{\pi}{2} - \tau - \theta_2 - \tau - \text{echo}$. Plain lines correspond to the
		fit of experimental data using eq.\eqref{eq:ID1}.}\label{fig:ID}
 \end{figure}

To observe the effect of ID on the electronic coherence time, we measured
the decay of the Hahn echo using eq. \eqref{eq:2PE_part2} and we varied the refocusing angle
$\theta_2$ using the sequence: $\frac{\pi}{2} - \tau - \theta_2 - \tau - \text{echo}$. Figure~\ref{fig:ID} 
shows the decay time $T^{-1}_2$ of \I at $T = 5.4$~K. The inset displays two examples
of Hahn echo decay with $\theta_2 = \pi/4$ (blue square) and $\theta_2 = \pi$ (orange circles). In
all our measurements using 2PE, the stretched parameter $n$  in eq.\eqref{eq:2PE_part2} was close
to 1 showing that ID plays an important role in the decay time. It is also important to mention that
the fits using eq.\eqref{eq:2PE_part1} were not satisfying since the relative weight of
$\Gamma_\text{SD}R$ was small compared to $\Gamma_{0}$. The decay times $T^{-1}_2$ of the 2PE
Hahn echo decay as a function of $\theta_2$ and $\left<\sin^2{\theta_2/2}\right>_f$ are presented
in Figure~\ref{fig:ID}(a) and (b) respectively. Using Eq.\eqref{eq:ID1}, we extract from the slope of Figure~
\ref{fig:ID}(b) $\pi\Gamma_{ID} = 294$ kHz, which leads to a concentration $C = 3.5 \times 10^{17}$
cm$^{-3}$. The values of $\pi\Gamma_{0,ID}$, $\pi\Gamma_{ID}$, and $C$ for \I, \Br, and \Cl are
provided in Table \ref{Tab:ID} and compared to values extracted from continuous wave (cw) ESR measurements
\cite{soriano2022,soriano2023}.

 \begin{table}[h] 
 	\caption{Fitted parameters of instantaneous diffusion ($\pi\Gamma_{0,ID}$, $\pi\Gamma_{ID}$) (see Fig. \ref{fig:ID} and details in the main text). 
 		$C_{ID}$ is the concentration extracted from eq.\eqref{eq:ID2} and $C_\text{cw}$ is the concentration obtained by cw-ESR measurements. The fit error estimation is included in the last digit.   \label{tab:impurity}}
 	\begin{center}
 		 \begin{tabular}{lcccc} \hline
 			\noalign{\smallskip}\hline
 			\noalign{\smallskip}
 			 &$\pi\Gamma_{0,ID}$&  $\pi\Gamma_{ID}$ & $C_{ID}$ & $C_\text{cw}$ \cite{soriano2022}  \\
 			 & (ms$^{-1}$) & (ms$^{-1}$) & $10^{17}$.cm$^{-3}$ & $10^{17}$.cm$^{-3}$\\
 			\noalign{\smallskip}\hline\noalign{\smallskip} 
 			\Cl& 460 & 236 & 2.8 & 11 \\
 			\Br& 304 & 309 & 3.7 & 13\\
 			\I & 208 & 294 & 3.5 & 8\\ 
 			\noalign{\smallskip}\hline\hline \end{tabular}\label{Tab:ID} \end{center}
 \end{table}

For all compounds, the concentration obtained from ID is significantly
lower than that determined by continuous wave electron spin resonance (cw-ESR). The concentrations
from cw-ESR were derived by comparing the relative intensities of the spin chain and end chain
signals, resulting in a very small error. In the case of ID, the fitting procedure is detailed in
the Supplementary Information (SI) (Fig. \ref{fig:SI__InsDiff_RAW_Br}, \ref{fig:SI__InsDiff_RAW_I},
\ref{fig:SI__InsDiff_RAW_Cl},\ref{fig:SI_ID_BrCl}), with a
cumulative error of less than 20\%. The measurement error alone cannot account for the discrepancy
in concentration resulting from the ID and cw-ESR studies. This discrepancy is a key finding of our work. For an ensemble of localized S=1/2 defects, the local spin concentration probed by dipolar interactions is expected to be equal than the average concentration ($C_{ID} = C_\text{cw}$). Our observation of $C_{ID} < C_\text{cw}$ strongly contradicts the localized spin model. It is important to note that the concentration
from cw-ESR is directly related to the number of S=1/2 spins, whereas the concentration from ID is
extracted from the effective dipolar field. The nature of the defect in quantum spin chains is
fundamentally different than that of isolated defects (such as metallic ions or free radicals). As explained
in the introduction and in Ref. \cite{eggert1995,martins1997,sorensen1998,hansen1999,nishino2000a}, the
spin S=1/2 we observed is made of dozens of coupled spins (see Figs. \ref{fig:localMag_DMRG} and \ref{fig:SI_localMag_DMRG2}). It
is not surprising that the strong correlation inside the spin chain might have an effect on the
external couplings. To support this hypothesis, we performed DMRG calculation on two QSC-ES
interacting though an Ising-like dipolar interaction
$d\boldsymbol{S}_{1}^z\cdot\boldsymbol{S}_{2}^z$. We then calculated the anistropic gap
$\Delta_\delta$ induced by $d$ for many values of the dimerization factor $\delta$  and compared
them with the anisotropic gap of 2 isolated spin $\Delta_{iso}$. The details of calculation are
presented in SI, and the main results are: For $\delta=1$ (\textit{i.e.} chain of non interacting dimers with a
isolated spin at the end), $\Delta_{\delta=1}=\Delta_{iso}$, and when $\delta$ decreases, the gap
$\Delta_\delta$ reduces and reach zero for $\delta=0$ (ie. the uniform spin chain). In the
particular cases of $\delta$ ranging from 0.08 to 0.096 for \DMTTFX \cite{soriano2022}, we obtain
$\Delta_{\delta=0.08}=0.29 \Delta_{iso}$ to $\Delta_{\delta=0.096}=0.35 \Delta_{iso}$ which match
with the reduction of the effective concentration observed by ID. Although this model is very
simple, it allows us to semi quantitatively explain the reduction in the effective dipole field
between 2 QSC-ES. This result can be interpreted as follows: QSC-ES are many-body objects and it is
therefore expected that the strong internal correlations arising from the isotropic J exchange
interaction reduce the effect of dipole interactions.

\subsection*{Spectral Diffusion}

\begin{figure}[h!]
	
	\includegraphics[width=\linewidth]{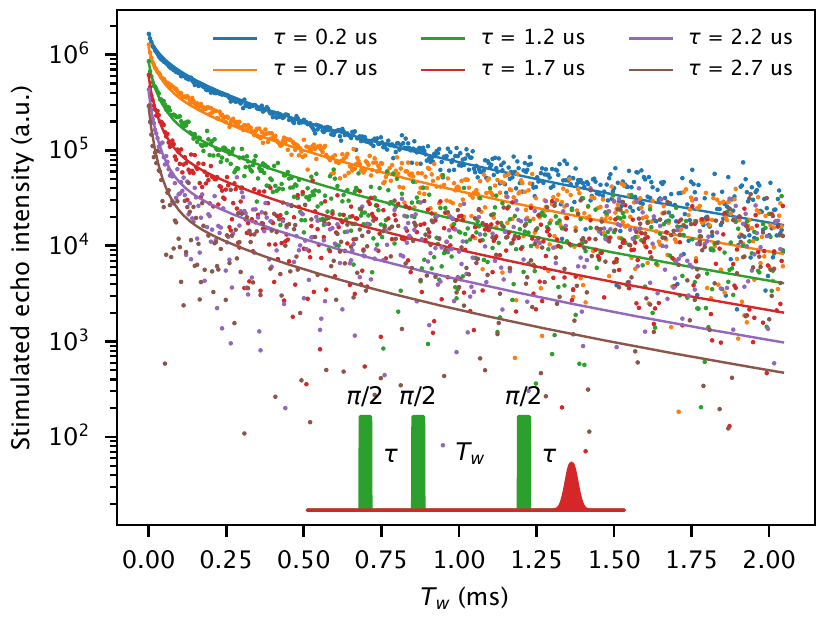}
	\caption{3PE measurements probing spin coherence dynamics in \I at 6~K.
		 The figure displays selected echo decay curves measured at various values of the delay time
	$T_W$. The decrease in signal intensity while increasing $T_W$ is attributed to spectral diffusion,
	homogeneous broadening and spin-lattice relaxation effects. The solid lines correspond to
	simultaneous fits of the entire set of 3PE data using equations \eqref{eq:I3p} and
	\eqref{eq:Gamma_eff}. The shared fit parameters  $\Gamma_0$, $\Gamma_{SD}$, $R$, $T_1$ and  $\beta$
	are reported in Table \ref{tab:3PE} for I, Br and Cl.} \label{fig:3PE}
\end{figure}

Now ID has been quantified in all \DMTTFX samples, we focus on  other sources of decoherence such as
homogeneous broadening and spectral diffusion (SD). 2PE and 3PE are complementary methods to extract
coherence contributions. 2PE cannot separate the linewidth of the spectral diffusion
$\Gamma_{\text{SD}}$ and the spin-flip rate $R$ (see. eq.\eqref{eq:2PE_part1}). Moreover, 2PE
sequence is accurate to extract spectral diffusion when the stretch exponent $n$ in eq.\eqref{eq:2PE_part2} 
is close to 2 while in our case it falls in the range 1 to 1.1. As a consequence,
we have first studied the 3PE sequence at the lowest temperature. We measured the decay of the
stimulated echo while increasing $T_W$ at a constant $\tau$ value and we have then varied $\tau$ over the range
of 200~ns to 4000~ns. This gives us a series of echo decays that we fitted simultaneously using
eq.\eqref{eq:I3p} and eq.\eqref{eq:Gamma_eff}. Selected stimulated echo decays for \I are presented in Figure \ref{fig:3PE}.

 \begin{table}[h] 
	\caption{Fitted parameters of spectral diffusion ($\pi\Gamma_{0}$, $\pi\Gamma_{SD}$, $R$, $T_1^{-1}$ and $\beta$) obtained from the stimulated echo
	decay measurements. For comparison,
	$T_{1,IR}^{-1}$ and $\beta_{IR}$ from inversion recovery measurements are also provided. The fit error estimation is included in the last digit.   }
	\begin{center}
		\begin{tabular}{lccccccc} \hline
			\noalign{\smallskip}\hline
			\noalign{\smallskip}
			$X$ &$\Gamma_0$&  $\Gamma_{SD}$ & R & $T^{-1}_1$& $\beta$ & $T^{-1}_{1,IR}$& $\beta_{IR}$  \\
			& (kHz) & (kHz) & (kHz) & (kHz)&  & (kHz) & \\
			\noalign{\smallskip}\hline\noalign{\smallskip} 
			Cl& 130 & 150 & 20 & 2.5 & 0.59 & 2.4 & 0.55\\
			Br& 110 & 120 & 27 & 2.0& 0.5 & 1.8 & 0.56\\
			I & 106 & 190 & 29.8 & 3.1& 0.56 & 3.2 & 0.68\\ 

			\noalign{\smallskip}\hline\hline \end{tabular}\label{tab:3PE} \end{center}
\end{table}

The extracted values of the spin-lattice relaxation $T^{-1}_1$ are coherent with the ones obtained
by inversion recovery $T^{-1}_{1,IR}$. The values of the stretched exponents obtained from 3PE (eq.
\eqref{eq:2PE_part1}) and inversion recovery (eq. \eqref{eq:fit_relaxation}) are comparable, showing
that the phonon-bottleneck process is visible also in 3PE. The values of $\Gamma_{SD}$ agree with
the dipolar interactions between QSC-ES. Values of the flipping rate $R$ are one order of magnitude
larger than what is expected for $T_1$ spin flipping only. If we take into account the flip-flop
process into the flipping rate (see \ref{SI:spinflip}) we obtain a total spin flip rate of
$R=24$~kHz close to our experimental value. Finally, we note that the homogeneous linewidth
$\Gamma_0$ is large compared to values reported in literature for equivalent spin concentration
\cite{bottger2006,rancic2022,lim2018}.

\begin{figure*}[t!]
	
	\includegraphics[width=\linewidth]{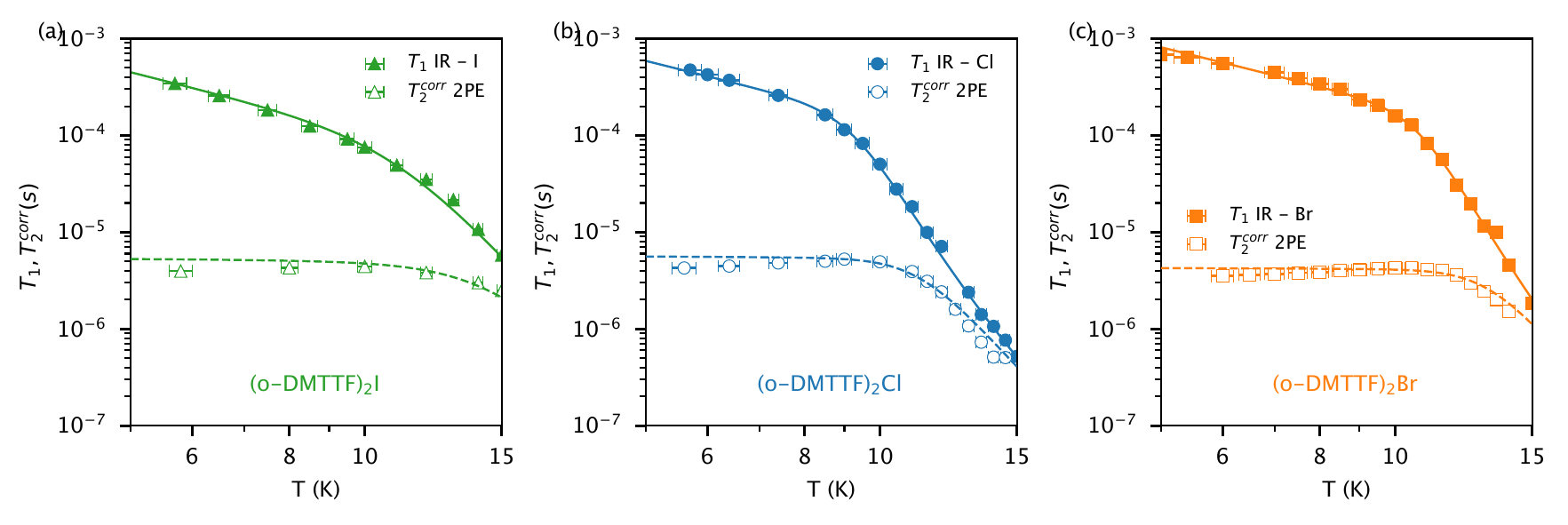}
	\caption{Temperature dependence of the corrected coherence time $T_{2-2PE}^{\text{corr}}$ (open
	symbols) and spin-lattice relaxation time $T_1$ (solid symbols) for \DMTTFX compounds (\(X =
	\text{Cl, Br, I}\)). The solid lines represent the phonon-bottleneck direct process and the
	high-temperature Raman or Orbach processes, as discussed in the main text. The dashed lines show the
	theoretical prediction of the coherence time using the parameters extracted from spectral diffusion
	analysis (see Table \ref{tab:3PE}), with no additional fitting parameters. The data demonstrate the
	interplay between spin-lattice relaxation and spectral diffusion in determining the coherence time
	of QSC-ES.}
	\label{fig:T1T2vsT}
\end{figure*}

Now, let's compare the spin diffusion parameters obtained by 3PE with the temperature dependence of the
2PE decay. For this study, we performed a 2PE decay as function of temperature and we extracted
the decoherence time $T_{2-2PE}$ using eq. \eqref{eq:2PE_part2}. We removed the ID contribution
using the relation:
$T^{-1\text{corr}}_{2-2PE}=T^{-1}_{2-2PE}-\pi\Gamma_{ID}\left<\sin^2{\pi/2}\right>_f$. Corrected 2PE
coherence time $T^{-1\text{corr}}_{2-2PE}$ and spin lattice relaxation time $T_1$ are provided in
Figure \ref{fig:T1T2vsT} for \I, \Cl and \Br. The solid lines are the relaxation time using phonon
bottleneck direct process and Raman or Orbach process as discussed previously. For the dashed lines we
used the definition of the coherence time \cite{klauder1962}:

\begin{equation} 
T_2=\frac{-a+\sqrt{a^2+2b/\pi}}{b}
\end{equation}

where $a=\Gamma_{0}$, and $b=\Gamma_{\text{SD}}R/2$ are taken from Table \ref{tab:3PE}. This
equation is valid for $T_2\ll T_1$. To take into account the high temperature regime, we also include
the limit due to spin lattice relaxation (see eq.\eqref{eq:T2GammaT1}). It is important to note
that no free fit parameters were used to simulate the dashed lines in Fig. \ref{fig:T1T2vsT}. In the
absence of fit parameters, our data describe accurately the temperature dependence of the 2PE.
However it appears that $T^{\text{corr}}_{2-2PE}$ is weakly sensitive to the diffusive part $\Gamma_{SD}R$ and
the decoherence is mainly due to the homogeneous linewidth $\Gamma_{0}$ which explains why we did not
extract accurately the spin diffusion contribution from 2PE.

The microscopic origin of $\Gamma_{0}$ remains an open question. The presence of nuclear spins in
\DMTTFX is not negligible, and in similar molecular magnets with an $S=1/2$ ground state, such as
V15 \cite{shim2012} and Cr7Ni \cite{kaminski2014}, they are the primary source of decoherence, often
exhibiting a significant ESEEM signature. However, in our study, no ESEEM signal from the systems was detectable, and
previous work has shown that the hyperfine interactions in QSC-ED are significantly reduced by the
exchange coupling within the spin chain \cite{soriano2020}.

Another possibility arises from the decoherence effect of the spins within the chain. Although we are probing
QSC-ED, which display a low concentration, they are composed of dozens of spins (see
Figs.~\ref{fig:localMag_DMRG}, \ref{fig:SI_localMag_DMRG2}) strongly correlated to the spin chain. In
such magnets, the homogeneous linewidth can be estimated using the theory of moments. Introduced by
Anderson and Weiss \cite{anderson1953} this method shows that the linewidth is proportional to
$\frac{\sum_i M_2^i}{J}$ where $M_2^i$ is the second order moment of the interaction involved in the
broadening such as electron-nuclear / electron-electron dipole interactions and anisotropic
super-exchange interaction. Although this method is fairly simplistic regarding of the structure of
QSC-ED, it does allow us to estimate the contribution of each interaction and their influence on the
linewidth. To estimate the electro-nuclear dipolar field, DFT calculations were performed on a single molecule of \I and have shown a
electro-nuclear dipolar field of  $M_2^{e-n}\sim 10^{13} \text{Hz}^2$ (see SI). The dipolar interaction caused by
all the spins inside the chains give $M_2^{e-e}\simeq 3\times 10^{17} \text{Hz}^2$  and anisotropic
exchange term although very small is estimated to $M_2^{ani}\simeq 3\times 10^{16} \text{Hz}^2$ (see
\ref{SI:moments}). Finally, using the estimation of $J\sim 600$~K obtained in
\cite{soriano2022,soriano2023}  we obtain an homogenous intrinsic linewidth $\Gamma_h\sim 25$~kHz
compatible with values reported in table \ref{tab:3PE}. Our method for estimating the sources of decoherence can be applied to other systems, in particular inorganic materials which were intensively studied in the 90s. As an example, let's confront the method to \ce{CuGeO3} doped Zn \cite{smirnov1998a}. While $M_2^{e-n}$ and $M_2^{e-e}$ remain weak, with $\Delta g\sim 0.1$ and $J\sim 100$~K we obtain $M_2^{ani}\simeq 5\times10^{20} \text{Hz}^2$ and an homogeneous coherence time of 1ns which explain why no coherence was reported in such material.  

\begin{figure}[t!]
	\includegraphics[scale=1]{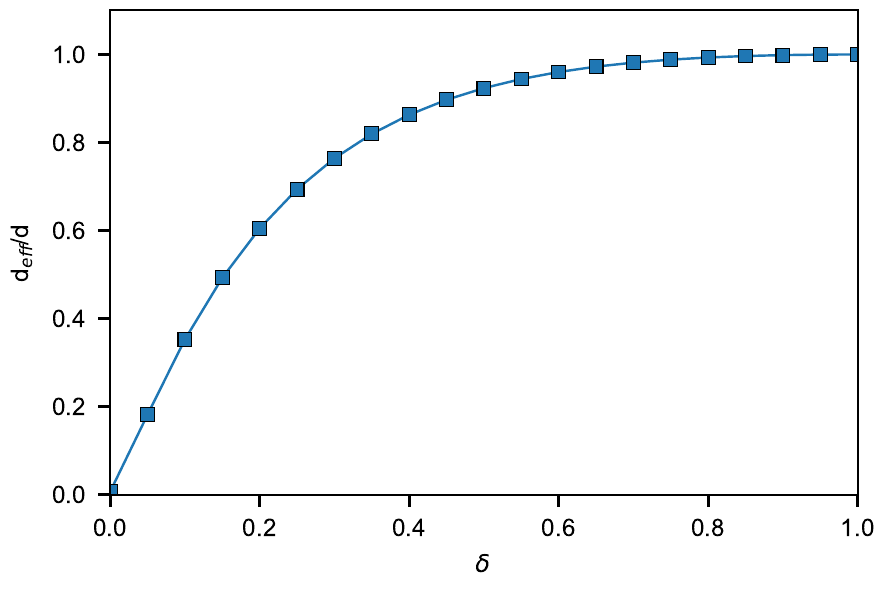}
	\caption{Effective dipole-dipole field as function of the dimerization factor $\delta$  }\label{fig:SI_DMRG_GAP}
\end{figure}

\add{\textit{Design principles for optimizing coherence.}---Our analysis provides actionable guidelines for engineering longer coherence times in quantum spin chain edge states. The key control parameter is the dimerization factor $\delta$: according to our DMRG calculations (Fig.~\ref{fig:SI_DMRG_GAP}), the effective dipolar coupling between edge states scales approximately as $d_{\rm eff}/d \approx 3\delta$ for small $\delta$. This suggests that materials with weaker dimerization will exhibit reduced instantaneous diffusion and longer $T_2$. However, reducing $\delta$ also increases the spatial extent of the edge state wavefunction (Fig.\ref{fig:localMag_DMRG}, \ref{fig:SI_localMag_DMRG2}), which may enhance homogeneous decoherence from intra-chain dipolar interactions. The optimal coherence is therefore expected at an intermediate $\delta$ value that balances these competing effects. For (o-DMTTF)$_2$X with $\delta \approx 0.08$--0.09, our data suggest that homogeneous broadening ($\Gamma_0 \sim 100$--130~kHz, Table~\ref{tab:3PE}) currently limits $T_2$, while instantaneous diffusion contributes comparably. To extend coherence beyond the microsecond regime achieved here, future materials design should target: (i) increased exchange coupling $J$ to further suppress the homogeneous linewidth via exchange narrowing, (ii) reduced proton density through deuteration to minimize residual nuclear contributions, and (iii) optimization of $\delta$ near 0.05--0.08 to balance spectral diffusion suppression against wavefunction delocalization.}

\section{Conclusion}

In summary, we have presented a comprehensive study of the quantum dynamics of quantum spin chain edge states using pulsed electron spin resonance. While pioneering ESR studies in the 1990s confirmed the existence of such topological states, \add{the mechanisms governing their coherent dynamics have remained unexplored until now}.

\add{Our key findings are threefold. First, the} spin-lattice relaxation of the edge state is intimately connected to the bulk properties of the spin chain: at high temperatures, relaxation proceeds via an Orbach process through the spin-Peierls gap $\Delta_{\rm SP}$ (when $\Delta_{\rm SP} < \theta_{\rm D}$), while at low temperatures a phonon-bottleneck regime dominates regardless of the counter-ion.

\add{Second,} we have identified and quantified all sources of decoherence. Our \add{central} result is that the dipolar fields---both electronic and nuclear---which \add{typically dominate decoherence in dilute paramagnetic systems,} are significantly renormalized by the intra-chain exchange coupling\add{. This reveals the intrinsically many-body nature of the QSC-ES}.

\add{Third,} DMRG calculations establish that the dimerisation factor $\delta$ is the key parameter controlling this renormalization: for large $\delta$ the edge state behaves as an isolated spin, while for small $\delta$ the effective inter-defect dipolar coupling is strongly suppressed. \add{This framework quantitatively explains the long-standing puzzle of absent spin echoes in the (TMTTF)$_2$X family ($\delta \sim 0.03$), where dipolar-driven spectral diffusion is so weak that only homogeneous decoherence remains.}

\add{Our results provide concrete guidelines for materials optimization: (i) smaller $\delta$ values lead to reduced instantaneous diffusion, (ii) larger exchange coupling $J$ reduces the homogeneous linewidth, and (iii) larger spin-Peierls gaps $\Delta_{\rm SP}$ extend $T_1$ at intermediate temperatures. We emphasize that demonstrating functional quantum devices based on these principles---such as two-qubit gates or quantum sensing protocols---remains an important challenge for future work, requiring advances in materials synthesis and nanoscale addressability.}

\add{Nevertheless, this study establishes the foundational understanding necessary for such developments. In particular, our} findings are directly relevant to the recent realization of topological edge states in \add{atomically precise nanographene structures \cite{fu2025,su2025,zhao2024}}, where the internal dipolar field \add{may constitute} the ultimate \add{limit to} quantum decoherence. \add{More broadly, this work demonstrates that many-body correlations in topological systems can provide an intrinsic protection mechanism against environmental decoherence---a principle that may extend to other classes of correlated quantum matter.}

\section*{Acknowledgments}
This work is supported by Agence Nationale de la Recherche (ANR project “DySCORDE”,
ANR-20-CE29-0011). Financial support from the IR INFRANALYTICS FR2054 for conducting the research is
gratefully acknowledged, A.M.K. is supported by ANR QuantEdu-France (22-CMAS-0001) and by France
2030 investment plan, as part of the Initiative d’Excellence d’Aix-Marseille Université – A*MIDEX
(AMX-22-RE-AB-199).
\bibliography{__Relax_references.bib}

\pagebreak
\clearpage
\onecolumngrid
\SI

\begin{center}
	\textbf{\large{\textit{Supplementary Information} \\\smallskip
			\bluetitle{Exploring electron spin dynamics in spin chains using defects as a quantum probe.}}}\\
\author{L.~Soriano}\email{loic.soriano@lncmi.cnrs.fr}\thanks{Present address: \affLNCMI }\affiliation{\affIMNP}
\author{G.~Berbaud}\affiliation{\affBIP}
\author{M.D.~Kuz'min}\affiliation{\affIMNP}
\author{H.~Vezin}\affiliation{\affLASIR}
\author{O.~Jeannin}\affiliation{\affISCR}
\author{M.~Orio}\affiliation{\affISM}
\author{M.~Fourmigu\'{e}}\affiliation{\affISCR}
\author{S.~Bertaina}\email{sylvain.bertaina@cnrs.fr}\affiliation{\affIMNP}
\end{center}\label{Supplementary}

\section{DMRG calculations}
The simulations of the quantum systems were carried out using tensor networks and the Density Matrix
Renormalisation Group (DMRG) method~\cite{white1992}. The tensor network calculations were performed
using the Python library \textsc{TeNPy}~\cite{hauschild2018a}. It is important to note that DMRG  is
an approximation of the true ground state. There can be potential issues with convergence and small
numerical errors, DMRG has shown is robustness for 1D non frustrated system. The SVD cut-off was set
to $10^{-10}$, while the convergence criterion for the ground energy was a relative change of
$10^{-10}$. We have calculated the ground state and sometimes the 1st excited state. Thanks to
\textsc{TeNPy}, the scripts run smoothly from some seconds to some minutes on the heavier
calculation on an Apple macbook pro 2021 M1Max with 64Go of RAM.

\begin{figure}[t!]
	\includegraphics[scale=.8]{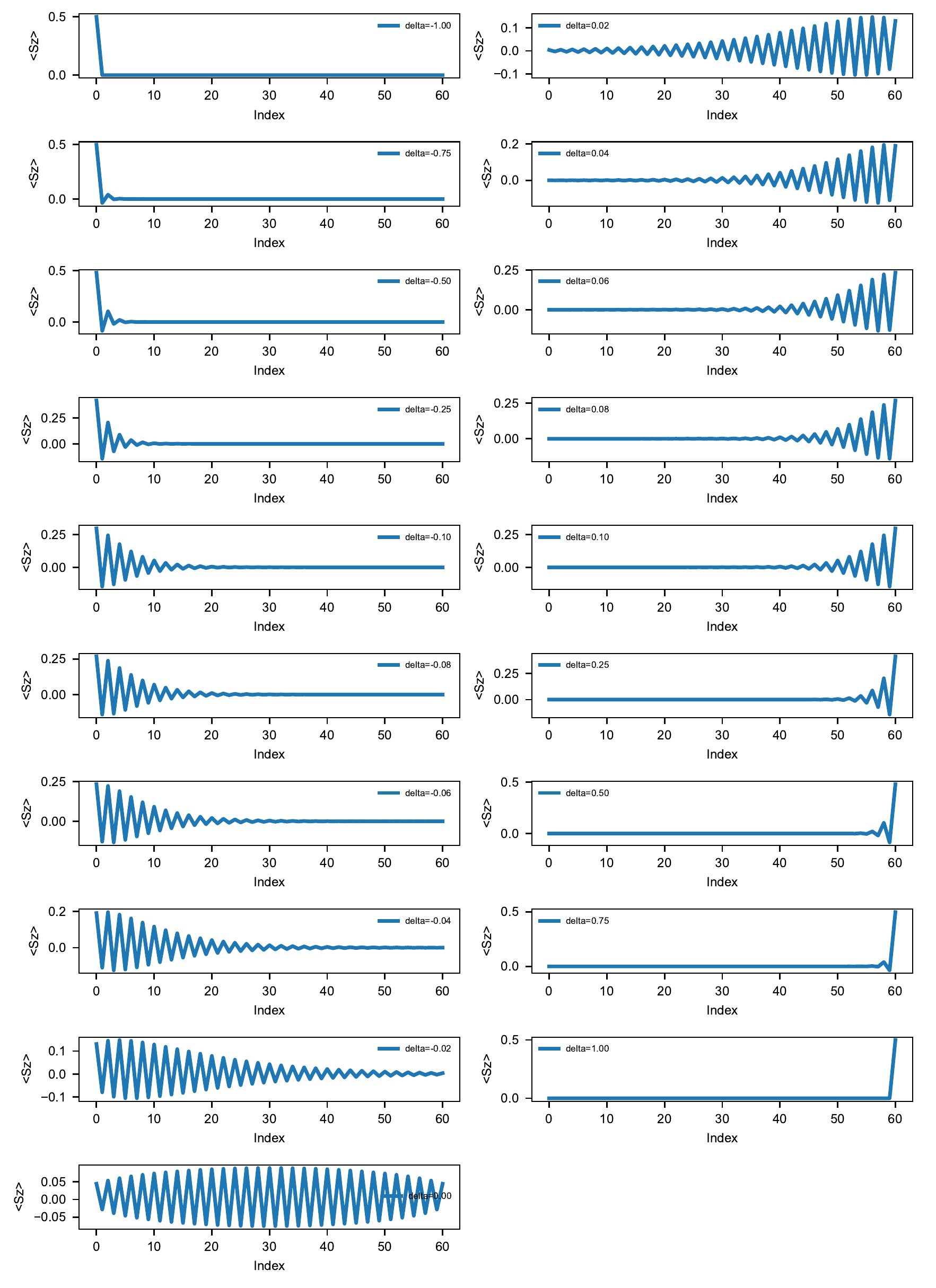}
	\caption{patial Magnetization Profiles for Open Spin Chains with Varying Dimerization. Local
	magnetization ($\langle S_z \rangle$) profiles, obtained through Density Matrix Renormalization
	Group (DMRG) calculations, are shown for open spin chains of $L$=60 sites. The dimerization
	parameter, $\delta$, is varied across a range spanning from -1.00 to +1.00, including $\delta$=0 to
	represent the uniform spin chain. For each value of $\delta$, the spin polarization arising from
	the open boundary conditions is clearly visible. These profiles illustrate the complex evolution of
	spin localization and correlation effects as the dimerization strength is modulated, transitioning
	from a regime of strong spin localization at the ends for large $|\delta|$ to a more homogeneous
	spin distribution for small $|\delta|$. Note that the edge polarization diminishes in magnitude as
	$|\delta|$ approaches zero.}\label{fig:SI_localMag_DMRG2}
\end{figure}

While changing the value of $\delta$ we have computed the distribution of magnetization for an odd
number of spin chain (see Fig. \ref{fig:SI_localMag_DMRG2}). It is worth to notice that for
$\delta=\pm 1$ which correspond to uncoupled dimer, the local magnetisation is zero excepted at the
site immediately after the defect. Wen decreasing $\|\delta\|$ the polarization spreads until
$\delta=0$ (\textit{i.e.} the uniform spin chain) where the magnetization is now through the entire chain.

In order to estimate the effect of the dimerization factor $\delta$ on the dipol-dipol interaction between two edge state, we have calculed the triplet gap between two 51 spins chains connected to an Ising interaction of value $d$. In the non interacting case: $\delta=1$ the coupling is only between the 2 isolated spin at the end. We can see that the dipol-dipol remains unchanged ($d_{eff}=d$). on the contrary in the case of uniform spin chain ($\delta$=0) the effective dipolar file tend to be extremely reduced ($d_{eff}\sim)0$. In the case of \DMTTFX, $\delta$ is close to 0.1 and $d_{eff}\sim 0.3$     

\newpage

\section{Spin-Phonon Interactions}
\label{spin_phonons}

To understand the processes involved in spin relaxation process, it is essential to distinguish
three subsystems: the spins, the lattice, and a thermal reservoir referred to as the bath
\cite{pescia1966}. The spins interact with lattice phonons via spin-orbit coupling: the lattice
vibrates. The lattice and the bath exchange heat through thermal contact. The lattice has a low heat
capacity, while the bath, by definition, has an infinite capacity. If the thermal contact between
the bath and the lattice is efficient, they remain in thermal equilibrium $T_p = T_0$. This
assumption is valid for describing the direct, Orbach, and Raman processes.

\begin{figure}\label{fig:SI_diagram}
	\centering
	\includegraphics[scale=1]{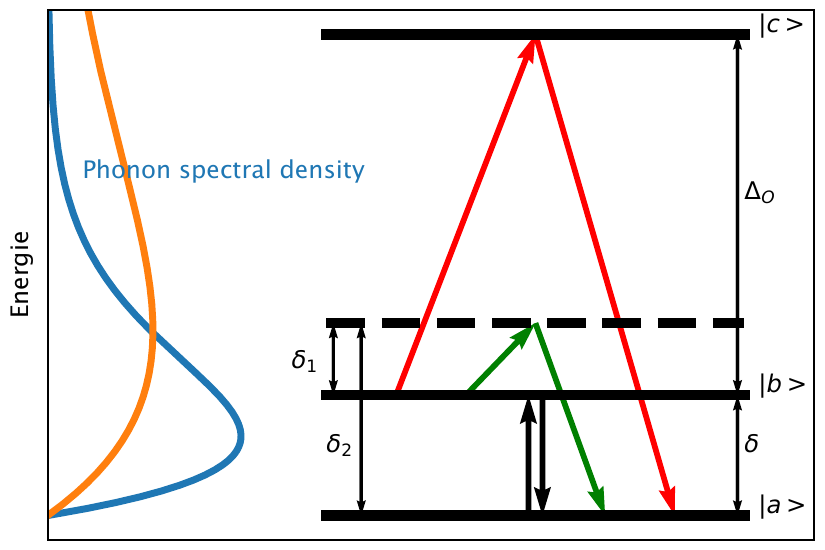}
	\caption{Diagram illustrating the three subsystems involved in lattice relaxation.}
\end{figure}

When the bath and the lattice exchange heat poorly, phonon-bath relaxation becomes non-negligible
compared to spin-phonon relaxation. The lattice phonons, being few in number, heat up to a
temperature $T_p \ne T_0$ and slow down spin-lattice relaxation. This phenomenon is called the
"phonon bottleneck" \cite{scott1962}.

\subsection{Direct Process}
\label{proc_direct}

Consider a two-level magnetic system $\{\ket{a}$,$\ket{b}\}$ separated by the Zeeman energy $E_z =
g\mu_B H_0$. The direct relaxation process results from the absorption and emission of a lattice
phonon with energy $\delta = \hbar\omega = g\mu_B H_0$. Here, it is assumed that the lattice and
bath are in continuous equilibrium. The spin populations $n_a$ and $n_b$ of each level evolve
according to equation (\ref{equilibre}), where $w_{a\rightarrow b}$ is the transition probability
from state $\ket{a}$ to state $\ket{b}$ per unit time, and vice versa for $w_{b\rightarrow a}$.

\begin{equation}
	\label{equilibre}
	\frac{dn_b}{dt} = -\frac{dn_a}{dt} = w_{a\rightarrow b}n_a - w_{b\rightarrow a} n_b
\end{equation}

The spectral density of the phonon population $\rho(\omega)$ is given by equation (\ref{ds_phonon}),
where $g_D(\omega)$ represents the Debye density of states in three-dimensional space, and $v$ is
the speed of sound in the medium \cite{abragam2013}. $K$ is a temperature-independent coefficient.

\begin{equation}
	\label{ds_phonon}
	\rho(\omega)d\omega = g_D(\omega)\bar{p}_0(\omega)d\omega
\end{equation}

\begin{equation}
	\label{pop_phonon}
	g_D(\omega) = \frac{3\omega^2}{2\pi^2v^3}\ ; \ \bar{p}_0(\omega) = \frac{1}{e^{\hbar\omega/k_BT_0}-1}
\end{equation}

\begin{equation}
	\label{proba_trans}
	w_{b\rightarrow a} = K[1+\bar{p}_0(\omega)] \ ; \ w_{a\rightarrow b} = K\bar{p}_0(\omega)
\end{equation}

Introducing $n = n_--n_+$ as the difference between the spin populations and $N = n_-+n_+$ as the total spin population, and combining equations (\ref{equilibre}) and (\ref{proba_trans}), we derive:

\begin{equation}
	\label{eq_IR}
	\frac{dn}{dt} = 2K[-n_+\bar{p_0}(\omega)-n_+ + n_-\bar{p_0}(\omega)]
\end{equation}

\begin{equation}
	\frac{dn}{dt} = K[(1+2\bar{p_0}(\omega)) n - N ]
\end{equation}

In equation (\ref{eq_IR}), we can identify, in sequence, stimulated emission, spontaneous emission,
and the absorption of a phonon with energy $\hbar\omega$. At equilibrium, $T_s = T_0$ and $ n = n_0
= N\tanh \left(\frac{\hbar\omega}{2k_BT_0}\right) $ is the thermal equilibrium polarization.

\begin{equation}
	\frac{dn}{dt} = -\frac{1}{T_{1d}}(n-n_0)
\end{equation}

From this, the temperature dependence of $T_1$ in the case of the direct process is deduced
(\ref{eq_direct}). The coefficient $K$ can be determined from $g(\omega)$ and by calculating the
transition probability using perturbation theory \cite{scott1962}. For a Kramers doublet, the
relation (\ref{eq_direct}) is found:

\begin{equation}
	\label{eq_direct}
	\frac{1}{T_{1d}} = \alpha_Dg^3B^5\coth\left(\frac{\hbar \omega}{2k_BT}\right)
\end{equation}

since our base temperature is 5.5K, $\hbar \omega \ll 2k_BT$ and we can approximate
$\coth\left(\frac{\hbar \omega}{2k_BT}\right)\approx \frac{2k_BT}{\hbar \omega}$ and so at the
resonance

\begin{equation}
	\label{eq_direct2}
	\frac{1}{T_{1d}} = \alpha'_DB^4_0 T
\end{equation}
for a Kramers doublet.

\subsection{Resonant Two-Phonon Process: Orbach}
\label{proc_orbach}
In the Orbach-type process, we consider a third energy level $\ket{c}$ separated by $\Delta_O$ from
the level $\ket{b}$ and by $\Delta_O + \delta$ from the fundamental level $\ket{a}$. This process
can only exist under the condition $\Delta_O < k_B\theta_D$, where $\theta_D$ is the Debye
temperature \cite{finn1961,orbachr.1961}. The transition from state $\ket{b}$ to $\ket{a}$ occurs in
two steps: absorption of a resonant phonon with energy $\Delta_O$ and emission of a resonant phonon
with energy $\Delta_O+\delta$, as shown in figure \ref{fig:SI_diagram}.

\begin{equation}
	\label{proba_trans1}
	w_{b\rightarrow c} = B[1+\bar{p}_0(\Delta_0)] \ ; \ w_{c\rightarrow b} = B\bar{p}_0(\Delta_0)
\end{equation}

\begin{equation}
	\label{proba_trans2}
	w_{a\rightarrow c} = B[1+\bar{p}_0(\Delta_0+\delta)] \ ; \ w_{c\rightarrow a} = B\bar{p}_0(\Delta_0+\delta)
\end{equation}

\begin{equation}
	\label{Orbach_IR}
	\left\{\begin{array}{cc}  \frac{d n_a}{dt} = B[n_c(\bar{p}_0(\Delta_0+\delta)+1)-n_a\bar{p}_0(\Delta_0+\delta)]\\
		\frac{d n_b}{dt} = B[n_c(\bar{p}_0(\Delta_0)+1)-n_b\bar{p}_0(\Delta_0)]
	\end{array}\right.
\end{equation}

In many cases, the spin population of the third energy level can be considered small compared to the
other two, $n_c \ll n_b, n_a$, and $\delta \ll \Delta_0$. The system of equations (\ref{Orbach_IR})
simplifies, and similarly to the direct process subsection \ref{proc_direct}, we find:

\begin{equation}
	\label{Orbach_IR1}
	\frac{dn}{dt} = -1/T_{1O}(n-n_0)
\end{equation}

It can be shown that for a Kramers doublet, $T_{1O}$ follows the relation (\ref{Orbach_T}):

\begin{equation}
	\label{Orbach_T}
	1/T_{1O} = \alpha_o\Delta_0^3e^{-\Delta_0/T}
\end{equation}

\subsection{Non-Resonant Two-Phonon Process: Raman}
\label{proc_raman}
The Raman process corresponds to the simultaneous absorption of a phonon with energy $\delta_1$ and
the emission of a phonon with energy $\delta_2 = \delta_1 + \delta$ to achieve a spin flip from
$\ket{b}$ to $\ket{a}$. Unlike the Orbach process, the phonons are not constrained by the value of
the gap $\Delta_O$, and if $\delta \ll k_B\theta_D$, then the entire phonon energy spectrum is
accessible up to $\theta_D$. The only condition to satisfy is $\delta_2 - \delta_1 = \delta$. To
calculate the transition probabilities per unit time $w_{b\rightarrow a}$ and $w_{a\rightarrow b}$,
we consider $H_c^{(2)}$, the second-order perturbation of the spin-phonon coupling Hamiltonian. We
then sum over all possible energies of $\delta_1$:

\begin{equation}
	\label{proba_Raman}
	w_{b\rightarrow a} = \int \frac{2\pi}{\hbar}|\bra{b,\underline{\bar{p_0}(\delta_1)},\bar{p_0}(\delta_2)}H_c^{(2)}\ket{a,\bar{p_0}(\delta_1)+1,\underline{\bar{p_0}(\delta_2)+1}}|^2\rho(\delta_2)\rho(\delta_1)d\delta_1
\end{equation}

\begin{equation}
	\label{proba_Raman2}
	w_{a\rightarrow b} = \int \frac{2\pi}{\hbar}|\bra{a,\underline{\bar{p_0}(\delta_1)+1},\bar{p_0}(\delta_2)+1}H_c^{(2)}\ket{b,\bar{p_0}(\delta_1),\underline{\bar{p_0}(\delta_2)}}|^2\rho(\delta_2)\rho(\delta_1)d\delta_1
\end{equation}

\begin{equation}
	\label{pop_Raman}
	\frac{dn_b}{dt}=-\frac{d n_a}{dt} = K' \int n_a\bar{p_0}(\delta_2)(1+\bar{p_0}(\delta_1)) - n_b\bar{p_0}(\delta_1)(1+\bar{p_0}(\delta_2)) \delta_1^6 d\delta_1
\end{equation}

Similarly to the direct process section \ref{proc_direct}, we find the equilibrium return equation (\ref{evolution_raman}):

\begin{equation}
	\label{evolution_raman}
	\frac{dn}{dt}=-\frac{1}{T_{1R}}(n-n_0)
\end{equation}

\begin{equation}
	\label{T1_raman}
	\frac{1}{T_{1R}} = K"\int_{0}^{\theta_D}\frac{\delta_1^6e^{\delta_1/k_BT}}{(e^{\delta_1/k_BT}-1)^2}
\end{equation}

The integral (\ref{T1_raman}) can be simplified if we assume that the temperature $T$ is low
compared to the Debye temperature $\theta_D$. This assumption is generally valid when measuring
$T_1$ at temperatures close to liquid helium. In the case where the system is a Kramers doublet, we
find the relation (\ref{T1_raman2}):

\begin{equation}
	\label{T1_raman2}
	1/T_{1R} = \alpha_RT^9
\end{equation}

The Raman process always exists, but the dependence of $T_{1R}$ on temperature or magnetic field is
strongly modified depending on the value of $\theta_D$, the nature, or the size of the system
\cite{gu2021}.

\subsection{Phonon Bottleneck Effect}
\label{proc_bottleneck}
The phonon bottleneck effect manifests when the thermal contact between the lattice and the bath is
poor. The lattice phonons are no longer necessarily in equilibrium with the bath at temperature
$T_0$ \cite{scott1962}. We now consider $\bar{p}(\omega)$ as the phonon population of the lattice at
temperature $T_p$.

\begin{equation}
	\label{pop_phonons}
	\bar{p}(\omega) = \frac{1}{e^{\hbar\omega/k_BT_p}-1}
\end{equation}

\begin{equation}
	\label{eq_pop}
	\frac{dn}{dt}=2\rho(\omega)[n_bw_{b\rightarrow a}(\bar{p}+1)-n_aw_{b\rightarrow a}\bar{p}]
\end{equation}

\begin{equation}
	\label{evolution_bottleneck}
	\left\{\begin{array}{cc} \frac{dn}{dt}=-\frac{1}{T_{1d}}[n-n_0-n\frac{\bar{p}-\bar{p_0}}{\bar{p_0}+1/2}] \\
		\frac{d \bar{p}}{dt} = -\frac{1}{\rho(\delta)\Delta\delta}\frac{dn}{dt}-\frac{\bar{p}-\bar{p_0}}{T_{ph}}
	\end{array}\right.
\end{equation}

Faughnan and Strandbergh numerically calculated the solutions (\ref{evolution_bottleneck})
\cite{faughnan1961} and showed that $n(t)$ and $\bar{p}(t)$ are not monoexponential. They
present a short relaxation time $T'_{1b}$ and a long relaxation time $T_{1b}$, both of which can be
expressed in terms of the bottleneck parameter $\sigma$.

\begin{equation}
	\label{time1}
	T_{1b} = (1+\sigma)T_{1d}
\end{equation}

\begin{equation}
	\label{time2}
	T'_{1b} = \frac{T_{ph}}{\sigma}
\end{equation}

The bottleneck factor $\sigma$ is very important. It is the ratio of the energy exchanged between the spins and the phonons relative to the energy exchanged between the phonons and the bath.

\begin{equation}
	\label{bottleneck_parameter}
	\sigma = \frac{E_{z}T_{ph}}{E_{ph}T_\text{SP}} = \frac{T_{ph}n_0}{T_{1d}\rho(\delta)\Delta\delta \coth{(\frac{\delta}{2k_BT})}}
\end{equation}

When $\sigma \ll 1$, the bottleneck effect is not visible, so $T_{1b} = T_{1d}$, and the direct process correctly describes the spin relaxation. Conversely, when $\sigma \gg 1$, we speak of a severe bottleneck effect, and the measured relaxation time is significantly extended, $T_{1b} \simeq \sigma T_{1d}$. The relation (\ref{severe_bottleneck}) is valid if $\delta \ll k_BT$:

\begin{equation}
	\label{severe_bottleneck}
	T_{1b} \simeq \sigma T_{1d} = D\coth(\frac{\delta}{2k_BT})^2 \simeq D'T^2
\end{equation}

\begin{equation}
	D' = \frac{6\Delta\delta k_B^2}{\pi^2n v^3 \hbar^3 T_{ph}}
\end{equation}

Faughnan and Strandbergh propose a simplified equation for $n(t)$ in the case $T_{1d} \gg T_{ph}$ (\ref{relaxation_bottleneck}) \cite{faughnan1961}. This condition is generally satisfied. The equation (\ref{relaxation_bottleneck}) has no trivial solution except for $\sigma = 0$, where $n(t)$ is an exponential law.

\begin{equation}
	\label{relaxation_bottleneck}
	-\frac{t}{T_{1d}} = (1+\sigma)\ln\left(\frac{n_0}{n-n_0}\right) -\sigma\frac{n}{n_0}
\end{equation}

The solution $n(t)$ can be numerically calculated from equation (\ref{relaxation_bottleneck}).
\subsection{Statistical Goodness of fit model}

In The main text, we present spin lattice relaxation of \Br, \Cl and \I using the chosen model (Orbach or Raman). Here we show the fit for both model confronted to every samples including error and $R^2$ estimation.    
\begin{figure}[t!]
	\includegraphics[scale=1]{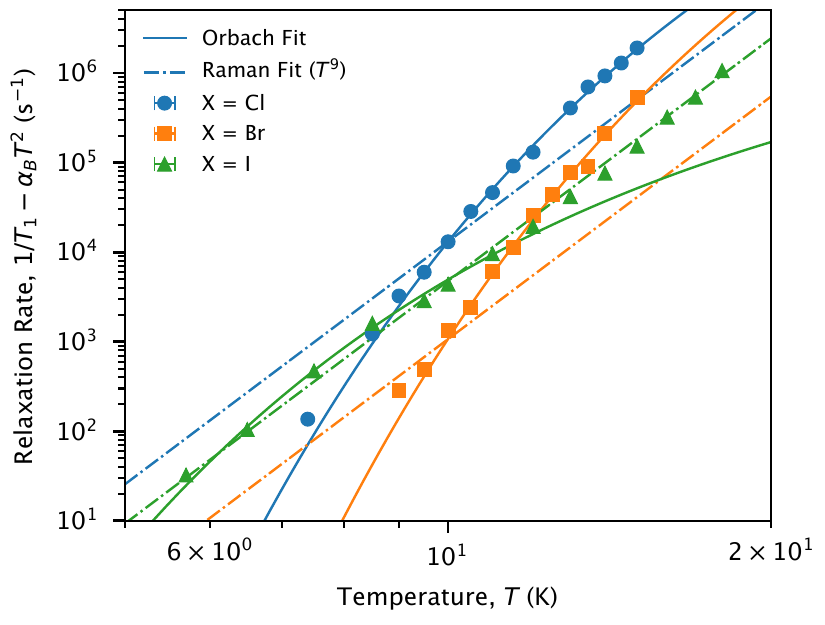}
	\caption{Spin lattice relaxation rate corrected by the phonon-bottleneck process presented in log-log scale. For the three compounds we have fitted using the two possible model : Orbach (plain line) and Raman (dashed dot line) processes }\label{fig:SI_T1_fit}
\end{figure}

 \begin{table*}[htb!] \caption{Fitted parameters of spin lattice relaxation.  \label{tab:SIT1}}
	
	\begin{widetable}{\textwidth}{lccccc} \hline
		\noalign{\smallskip}\hline
		\noalign{\smallskip}
		& Model &   $\alpha_{R9}$  &  $\alpha_O$ &  $\Delta_O$ & $R^2$ \\
		&  &  $\times 10^{-6}(\text{s}^{-1}\text{K}^{-9}$) &  $\times 10^{4}(\text{s}^{-1}\text{K}^{-3})$& (K)&\\
		\noalign{\smallskip}\hline\noalign{\smallskip} 
		\text{\Cl}& Raman & 13.1 $\pm$ 3.3 & - & - & 0.5085\\
		& Orbach & - & 1.13 $\pm$ 0.3 & 147 $\pm$ 4 & 0.9961\\
		
		\text{\Br}	& Raman & 1.1$\pm$ 0.8 & - & - & 0.201\\
		& Orbach & - & 1.6$\pm$ 0.4 & 181$\pm$17 & 0.9851\\

		\text{\I} & Raman & 4.17$\pm$ 0.17 & - & - & 0.9889\\ 
		& Orbach & - & 0.0015$\pm$0.0001 & 70 $\pm$ 2& 0.4474\\

		\noalign{\smallskip}\hline\hline 
	\end{widetable}\label{Tab:SIT1} 
\end{table*}
The value of $R^2$ clearly prove that Orbach process is involved in the relaxation of \Cl and \Br while Raman process dominates in \I.

\subsection{Spin lattice relaxation of \Cl and \I}

The main text presents the spin lattice relaxation data of \Br. Here we show the remaining systems.


\FloatBarrier
\pagebreak

\section{Instantaneous diffusion}

\begin{figure}[ht]
	\includegraphics[scale=.9]{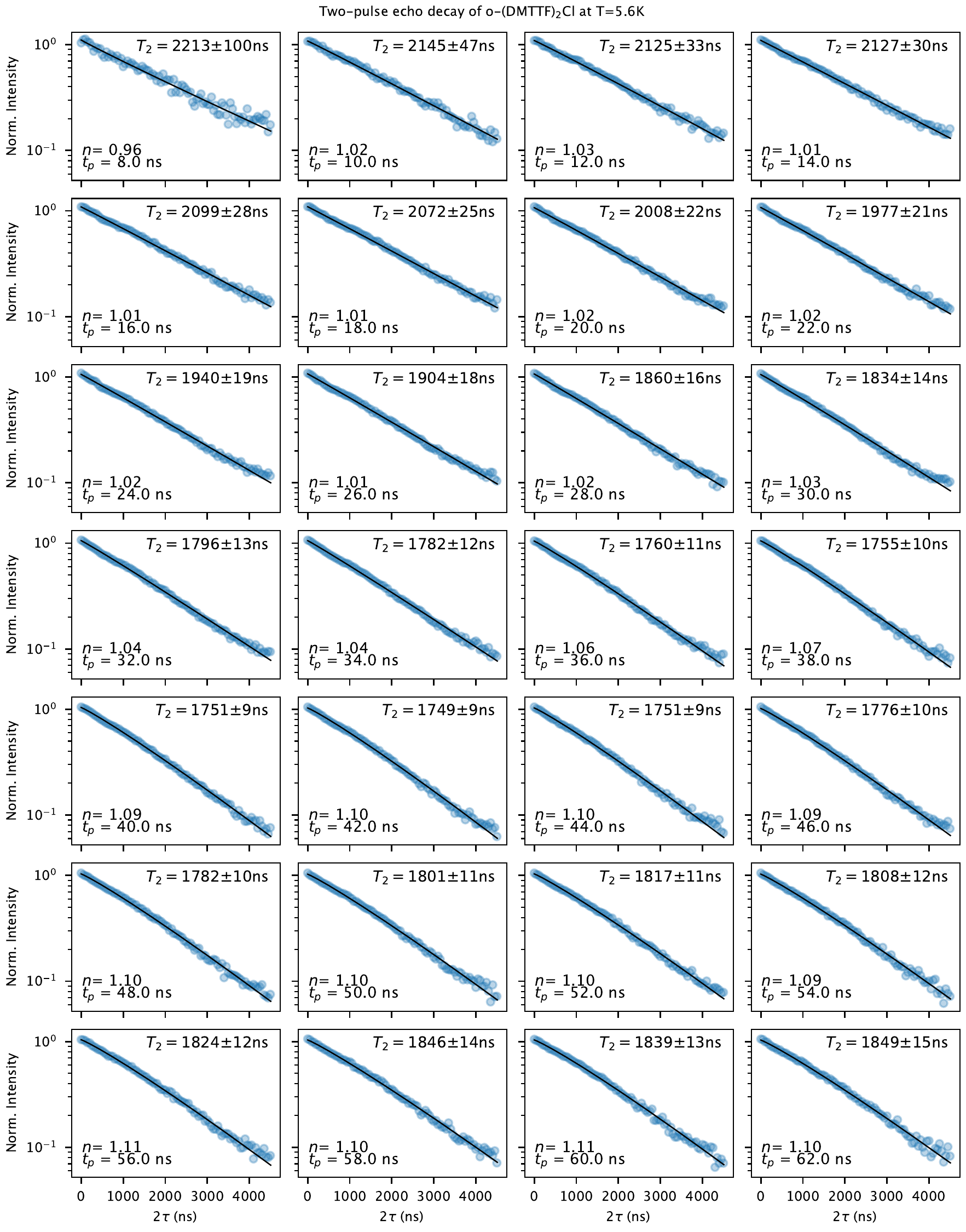}
	\caption{Two-pulse echo decay measurements for \Cl at T = 5.6K. The series of plots shows the decay
		of the spin echo signal as a function of the delay time $\tau$ between the excitation pulse
		($\pi/2$) and the refocusing pulse ($\theta_2$) in duced by different durations (tp) of the second
		pulse. These pulse durations are indicated in each graph, showing a change of decoherence rate for
		different duration pulse. Each solid blue lines corresponds to single exponential fits, resulting
		in the determination of T2 values displayed at the top of each panel}\label{fig:SI__InsDiff_RAW_Cl}
\end{figure}

\begin{figure}[th!]
	\includegraphics[scale=1]{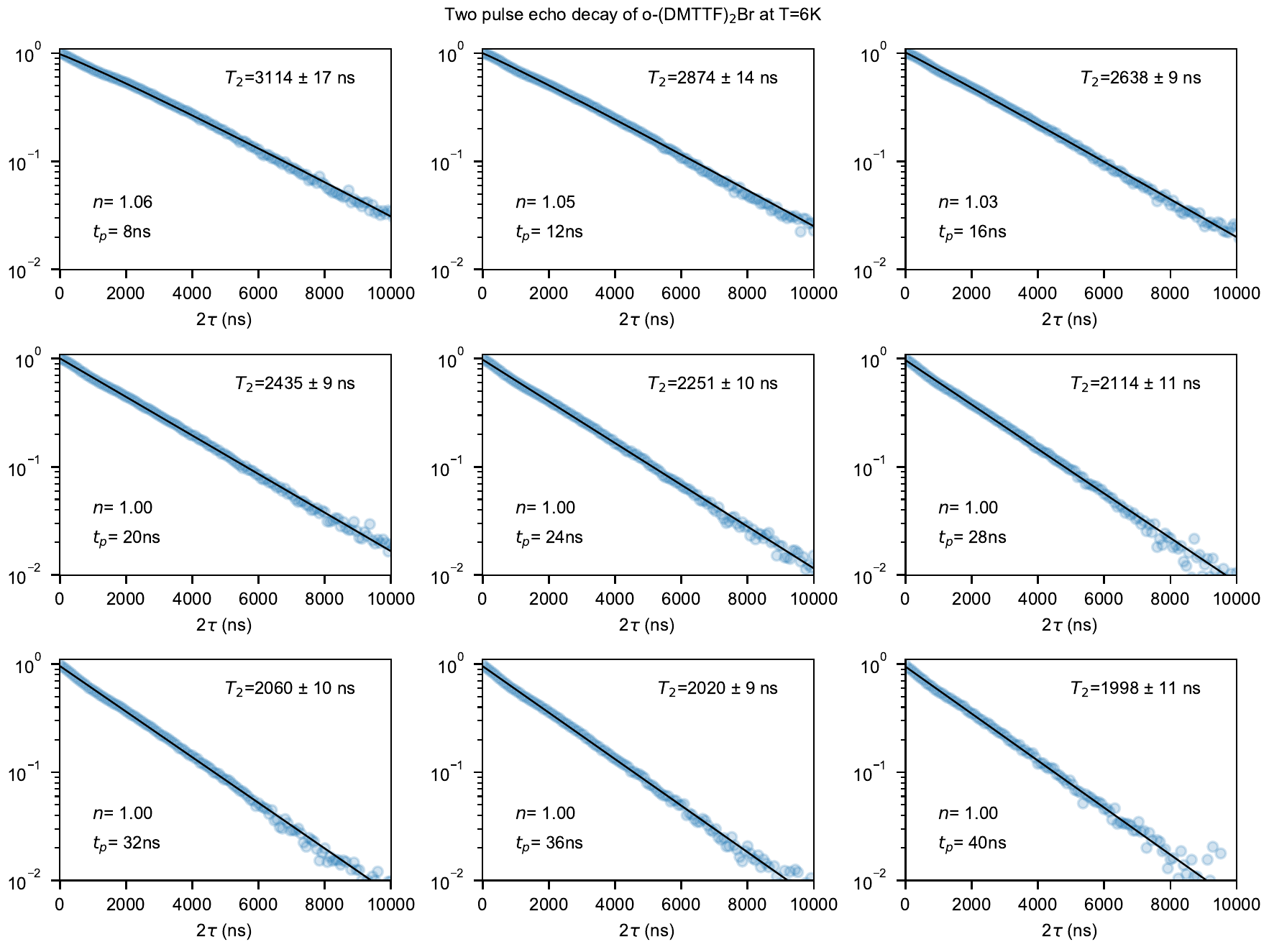}
	\caption{Two-pulse echo decay measurements for \Br at T = 6K. The series of plots shows the decay
		of the spin echo signal as a function of the delay time $\tau$ between the excitation pulse
		($\pi/2$) and the refocusing pulse ($\theta_2$) in duced by different durations (tp) of the second
		pulse. These pulse durations are indicated in each graph, showing a change of decoherence rate for
		different duration pulse. Each solid blue lines corresponds to single exponential fits, resulting
		in the determination of T2 values displayed at the top of each panel}\label{fig:SI__InsDiff_RAW_Br}
\end{figure}

\begin{figure}[t!]
	\includegraphics[scale=.9]{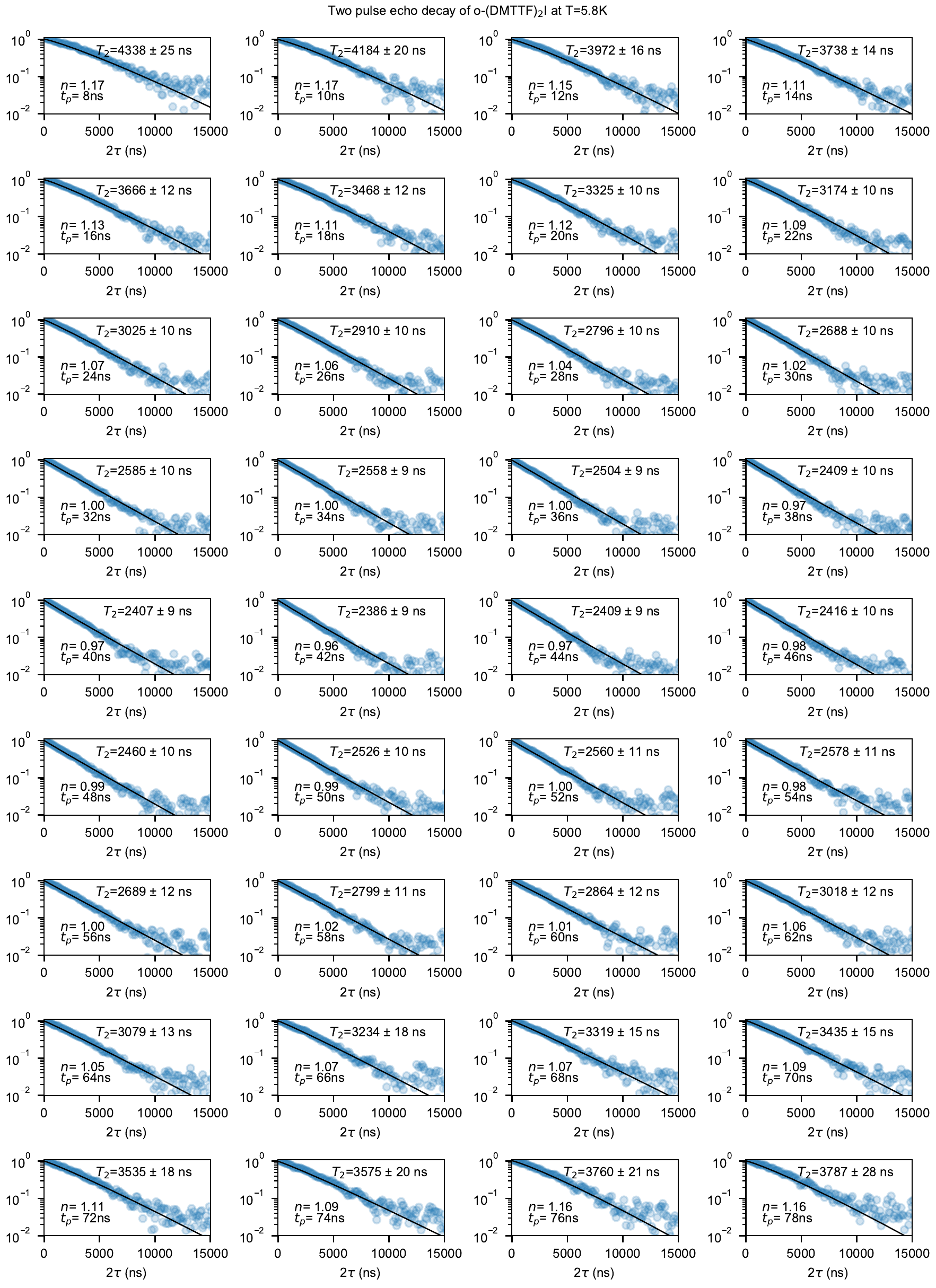}
	\caption{Two-Pulse Echo Decay Measurements for \I at 5.8 K with Varying Pulse Durations. A series
	of two-pulse Hahn echo decay curves are presented for \I, measured at a temperature of 5.8 K. Each
	panel shows the decay of the spin echo signal amplitude as a function of the delay time (2$\tau$)
	between the excitation pulse ($\pi$/2) and the refocusing pulse $\theta_2$. The duration of the
	refocusing pulse (denoted as $t_{p}$), is systematically varied for each decay series. The solid
	lines represent the best fits to the experimental data using a stretched exponential decay
	function, from which the coherence time (T$_{2}$) and the stretched parameter $n$ is extracted.
}\label{fig:SI__InsDiff_RAW_I}
\end{figure}

\begin{figure}[t!]
	\includegraphics[scale=.9]{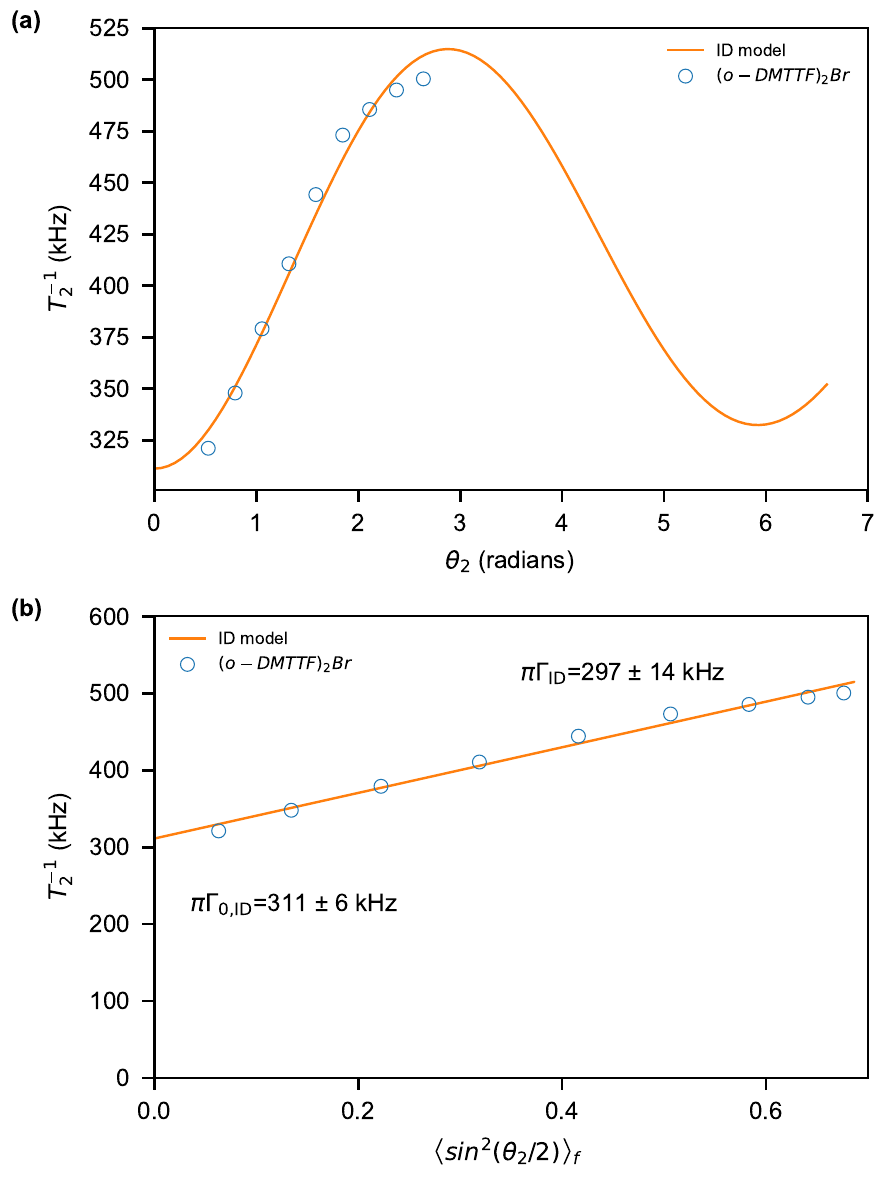}
	\includegraphics[scale=.9]{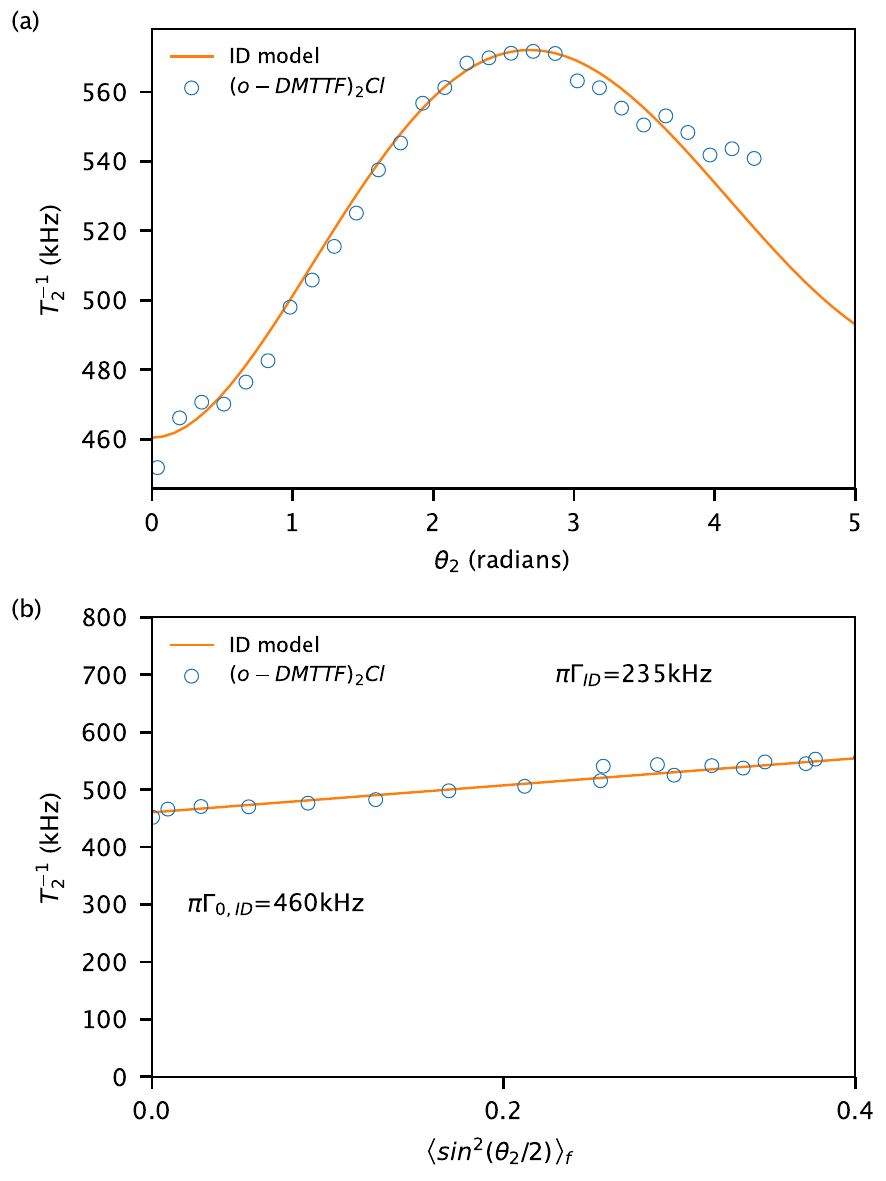}
	\caption{Two-Pulse Echo Decay Measurements for \Br, and \Cl Compounds. This figure presents a
	series of two-pulse echo decay curves using the same protocol than Fig. \ref{fig:ID}. Each plot
	illustrates the decay of the spin echo signal amplitude as a function of the delay time (2$\tau$)
	between the excitation pulse ($\pi/2$) and the refocusing pulse ($\theta_2$). The refocusing pulse
	duration ($t_p$) is systematically varied to observe changes in the decoherence rate. Solid lines
	represent the best fits to the experimental data using \eqref{eq:ID1}}
	\label{fig:SI_ID_BrCl}
\end{figure}

The main text introduces instantaneous diffusion as a significant decoherence mechanism affecting the Hahn echo decay. It outlines the theoretical basis where the coherence time $T_2$ is shown to depend on the refocusing pulse angle ($\theta_2$) due to ID. 
While the main text features a detailed analysis of ID for \I (Fig. \ref{fig:ID}), this section provides the equivalent experimental two-pulse echo (2PE) decay measurements and the fitting according to Eq. \eqref{eq:ID2} for the \Cl and \Br systems (shown in Figures \ref{fig:SI__InsDiff_RAW_Cl} to \ref{fig:SI_ID_BrCl}). Figures \ref{fig:SI__InsDiff_RAW_Cl}, \ref{fig:SI__InsDiff_RAW_Br} and \ref{fig:SI__InsDiff_RAW_I} display the raw 2PE decays for various refocusing pulse durations ($t_p$, which is proportional to $\theta_2$), from which individual $T_2$ values are extracted. Figure \ref{fig:SI_ID_BrCl}) then shows the plots of $1/T_2$ versus the refocusing pulse parameters for chloride and bromide, analogous to Figure \ref{fig:ID} for the iodide compound. In our experiment $\pi/2=24$ ns and $\theta_2$ was set by changing the pulse time from 8ns every 2ns
	
\subsubsection*{Concentration calculation procedure}
In the main text we compare two concentration values for each compounds. One obtained from a previous work \cite{soriano2022} by continuous wave EPR ($C_\text{cw}$)and one obtained from instantaneous diffusion $C_\text{ID}$ (present work). Here was details the procedures: 
We start with the concentration in "defects per formula unit" from \cite{soriano2022}, which was determined by fitting the intensity of the EPR line and is directly related to the static susceptibility.  We convert this dimensionless value to a volumetric concentration (in cm$^{-3}$) using the molar mass of the \DMTTFX compound and its crystallographic density, which is known accurately from X-ray diffraction data (\cite{foury-leylekian2011}). The formula used is: ($C_\text{cw}$ = (defects/formula unit)
 $\times$ ($\rho \times N_A) / M$, where $\rho$ is the density, $N_A$ is Avogadro's number, and $M$ is the molar mass. The main source of error comes from the fit of the EPR susceptibility data, which we estimated to be around 10\% is due to the error in quantitative EPR. This estimation of error is classic in the quantitative EPR of isotropic S=1/2 spins.

Regarding $C_\text{ID}$ obtained by instantaneous diffusion, the method is completely different. We measured the echo decay by increasing the refoculisation pulse length, then we fitted the decay time and the standard deviation error which is usually within 1\% error. We then plot the inverse of the decay time as function of $<\sin^2\theta>$ and we have fitted the slope with an error of about 4\%.

\begin{figure}[t!]
	\includegraphics[scale=1]{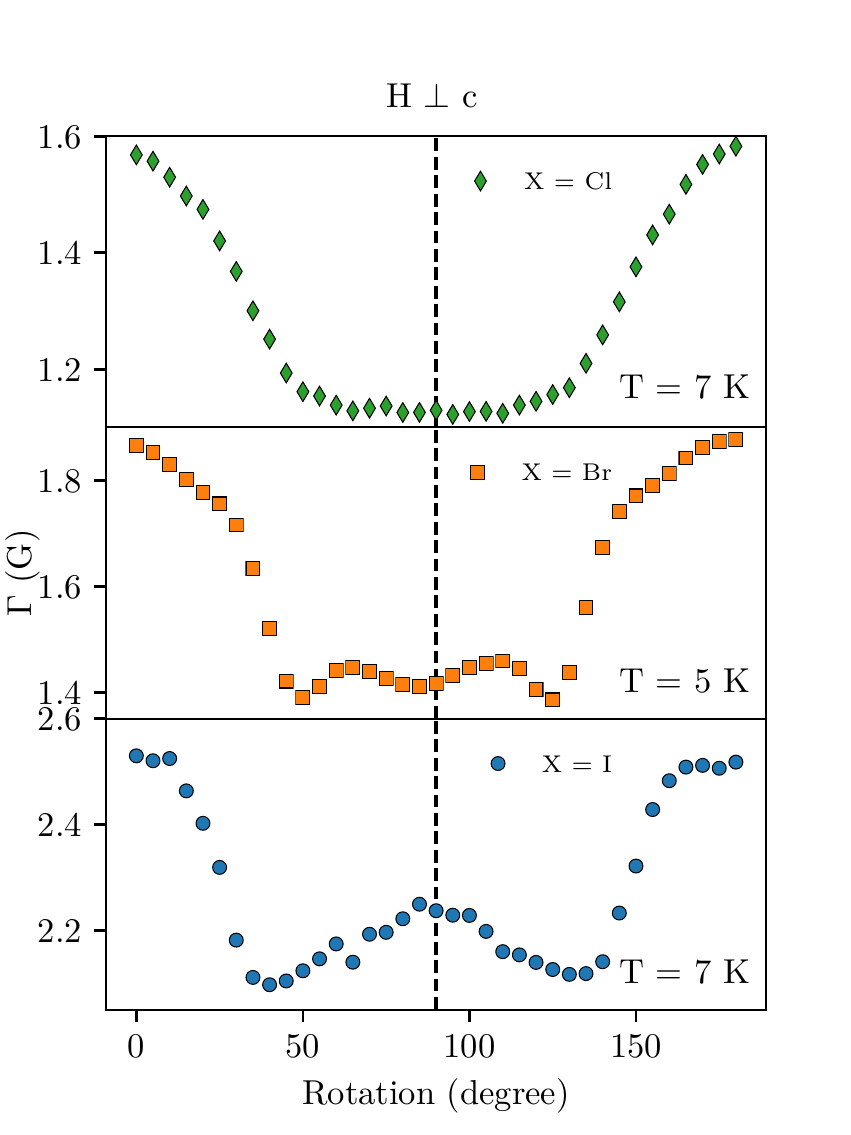}
	\caption{Angular dependence of the half width at half maximum with respect of the external maagnetic field. The purpose of this fugure is to provide the linewidth used for the ID calculation.}\label{fig:linewidth}
\end{figure}
\FloatBarrier
\pagebreak

\section{Spin-Flip Rate}\label{SI:spinflip}
The spectral diffusion analysis reveals the crucial role of the spin-flip rate, denoted by R, in
governing electron spin decoherence. This rate quantifies the frequency with which a spin in the
surrounding environment, either electronic or nuclear, undergoes a transition that affects the local
magnetic field experienced by the central, probed spin. The spin-flip rate, R, of group B spins acts
as a key parameter within our theoretical model, describing the influence of the surrounding spin
environment on the probed electron spin. Physically, this rate reflects the frequency at which spins
in the environment undergo transitions, such as spin-lattice relaxation or flip-flops, thereby
changing the local magnetic field and contributing to decoherence of the central spin. Our analysis
of the spin-flip rate assumes, for simplicity, an isotropic g-factor. In this case, the spin flip
rate can be decomposed into contributions from two distinct processes: spin-lattice relaxation (with
rate $1/T_1$) and flip-flops (with rate $R_\text{ff}$). The spin-lattice relaxation process is
mediated by phonons and is characterized by the interaction of the spin with the lattice vibrations,
and is described by the Fermi Golden Rule. The flip-flop process, on the other hand, involves the
mutual exchange of spin states between two or more interacting spins.

\subsubsection{Spin-lattice flip rate}
As explained in Section \ref{sec:SLR}, relaxation is dominated by the direct phonon bottleneck
process in our temperature range. If the perturbing spins are Kramers ions, their $T_1$ spin flip
rate is
  \begin{equation}
	\frac{1}{T_1^{\mathrm{phb}}} = \frac{3 \omega^{2} \Delta \omega}{2 \pi^{2} v^{3} n \tau_{\mathrm{ph}}} \operatorname{coth}^{2}\left(\frac{\hbar \omega}{2 k_{\mathrm{B}} T}\right)
\end{equation}

where $v$ the velocity of sound in the crystal, $\Delta \omega$ is the resonance linewidth, $n$ is
the concentration of spins, and $\tau_{\mathrm{ph}}$ is the lattice–bath relaxation time. Assuming
$k_BT \gg \hbar\omega$, and the inhomogenous linewidth is field dependent we can simplify by

\begin{equation}
	\frac{1}{T_1^{\mathrm{phb}}} = \alpha'_B H_0 T^2
\end{equation}

\subsubsection{Flip-flop rate}
The flip-flop rate can be estimated following the analysis done in \cite{bottger2006}:
      
\begin{equation}
	R_\text{ff} \approx \frac{2\pi}{\hbar} \left( \left|\left<\uparrow\downarrow | H_{dd, spin\, i-spin \, j} | \downarrow\uparrow \right>\right|^{2}\right)_{\text{avg}} \frac{1}{\hbar \Gamma_{inh}} = \left(\frac{\pi\mu_0\mu_B g^2C  }{9\sqrt{3}\hbar}\right)^2.\frac{1}{\Gamma_{inh}}
\end{equation}
Using the concentration of QSD-EC and the linewidth of figure \ref{fig:linewidth} we obtain :

$R_\text{ff}=$ 16.5~kHz, 24.3~kHz and 14~kHz for \Cl, \Br and \I respectively.

\section{Estimation of internal process in the decoherence}\label{SI:moments}
\subsection{Method of moments}
In the case of pure random process, Anderson and Weiss \cite{anderson1953} and Kubo et Tomita
\cite{kubo1954} have estimated that the sources of broadening which are included in the second
moment of the distribution is reduced by the exchange interaction J.

The method of moments is a statistical technique used to analyze the shape and characteristics of
spectral lines, particularly when the line shape is described by a Gaussian function. A Gaussian
function is given by:
\[
f(x) = \frac{1}{\sigma \sqrt{2\pi}} e^{-\frac{(x - \mu)^2}{2\sigma^2}}
\]
where \( \mu \) is the mean and \( \sigma \) is the standard deviation.

For a distribution $f$, the \( n \)-th moment about the mean is defined as:
\[
M_n = \int_{-\infty}^{\infty} (x - \mu)^n f(x) \, dx
\]

The second moment about the mean, measures the spread or width of the Gaussian spectral line:
\begin{equation}
	M_2 = \int_{-\infty}^{\infty} (x - \mu)^2 f(x) \, dx = \sigma^2
\end{equation}
Considering the half width at half maximum of the distribution $\Gamma_d$, we obtain
$M_2=\Gamma^2/2\ln2$.

For electron electron interaction we obtain : $M_2^{e-e}=\left(\mu_0\mu_B^2g^2
C_{tot}/(9\sqrt(3)\hbar\right)^2/2\ln2\simeq 3\times 10^{17} \text{Hz}^2$ where $C_{tot}$ is the
total concentration of spins \\

For anisotropic exchange interaction we used $M_2^{ani}=\left((\Delta g/g)^2J\right)^2$, using
$\Delta g / g= 0.004$ and $J\sim 600$~K \cite{soriano2022} we obtain  $M_2^{ani}=4 \times 10^{16}
\text{Hz}^2$

\section{Stretched exponential}

\begin{figure}[h!]
	\includegraphics[scale=1]{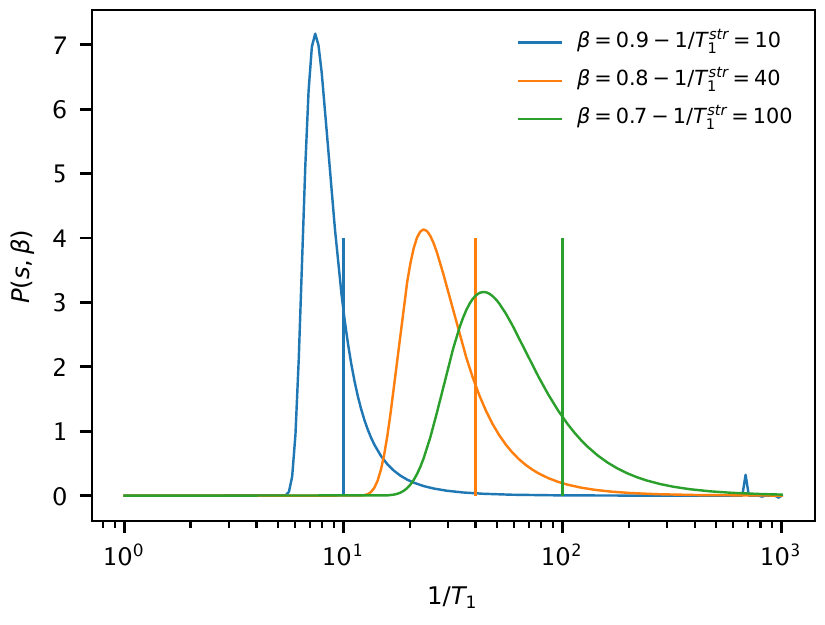}
	\caption{ The probability density function, $P(s, \beta)$, which represents the distribution of relaxation rates ($s$), is shown for various stretched parameters: $\beta = 0.9$ with $1/T_{\text{str}} = 10$ s$^{-1}$; $\beta = 0.8$ with $1/T_{\text{str}} = 40$ s$^{-1}$; and $\beta = 0.7$ with $1/T_{\text{str}} = 100$ s$^{-1}$. 
		$P(s, \beta)$ was computed using Equation (\eqref{eq:real_part}), illustrating how different values of $\beta$ and $1/T_{\text{str}}$ correspond to different distributions. A vertical line marks the input value of $1/T_{\text{str}}$ for each set of parameters to guide the eye.  }\label{fig:SI_relacdistrib}
\end{figure}

The stretched exponential function arises in systems where relaxation dynamics are influenced by a
distribution of rates instead of a single relaxation process. Following Johnston article
\cite{johnston2006} let's summarize the main results. Experimentally we can fit the spin lattice
relaxation using :
\begin{equation}
	\langle S_z \rangle = \langle S_z \rangle_0 \exp \left[-\left(\frac{t}{T_1^{\text{str}}}\right)^{\beta}\right]
	\label{eq:stretched_exponential}
\end{equation}
in the case of $0<\beta<1$ eq. \eqref{eq:stretched_exponential} becomes

\begin{equation}
	\exp \left[-\left(\frac{t}{T_1^{\text{str}}}\right)^{\beta}\right] = \int_{0}^{\infty} P(s, \beta) e^{-s \frac{t}{T_1^{\text{str}}}} \, ds
	\label{eq:integral_representation}
\end{equation}
where $s=\frac{T_1^{\text{str}}}{T_1}$ and  $P(s, \beta)$ the probability density of $s$ knowing
$\beta$. In the particular case of $\beta=1$, the relaxation is a monoexponential decay, $P(s,
\beta=1)$ is a Dirac function $\delta(s=1)$ and $T_1^{\text{str}}=T_1$. For a general $\beta$, eq.
\eqref{eq:integral_representation} is Laplace transform of  $P(s, \beta)$ and consequently, $P(s,
\beta)$ can be obtained by the inverse Laplace transform of the stretched exponential:

\begin{equation}
	P(s, \beta) = \frac{1}{2 \pi i} \int_{-i \infty}^{i \infty} e^{-x^{\beta}} e^{s x} \, dx
	\label{eq:inverse_laplace}
\end{equation}
Changing  $u=-ix$, Johnston shown that 

\begin{equation}
	P(s, \beta) = \frac{1}{\pi} \int_{0}^{\infty} e^{-u^{\beta} \cos (\pi \beta / 2)} \cos \left[s u - u^{\beta} \sin (\pi \beta / 2)\right] \, du
	\label{eq:real_part}
\end{equation}

which can be evaluated numerically. Some examples are provided in Fig.  \ref{fig:SI_relacdistrib}.
It is important to notice that the maximum of the distribution is not directly related to the data
extracted relaxation time $T_1^{\text{str}}$. However, Johnston has shown that that \(
\frac{1}{T_1^{\text{str}}} \) is not the average relaxation rate but rather a characteristic
property of the probability distribution \( P(s, \beta) \). Specifically, \(
\frac{1}{T_1^{\text{str}}} \) is the value such that relaxation rates \( \frac{1}{T_1} \) are about
equally likely to be less than \( \frac{1}{T_1^{\text{str}}} \) as they are to be greater than \(
\frac{1}{T_1^{\text{str}}} \). This means \( \frac{1}{T_1^{\text{str}}} \) can be interpreted as a
median relaxation rate within the distribution of relaxation rates.

\section*{DFT calculations}

\subsection*{ Computational details}
Theoretical calculations were based on the Density Functional Theory (DFT) and were performed using the ORCA 6.0 program.\textsuperscript{[1,2]} To facilitate comparison between theory and experiment, the X-ray crystal structure of (o-DMTTF)$_2/$ was used. The DFT model includes two dimethetyltetrathiafulvalene units together with 6 bromine counter-ions. The hydrogen atoms of the DFT model were geometry-optimized while constraining the positions of other atoms to their experimentally derived coordinates. All calculations were carried out using the hybrid functional B3LYP\textsuperscript{[3,4]} and the x2c-TZVPall basis sets.\textsuperscript{[5]} Relativistic effects were included using the exact two-component (X2C) Hamiltonian.\textsuperscript{[6-8]} Tight convergence criteria were imposed using the TightSCF keyword along with the default (DefGrid2) integration grids. Dispersion correction was applied according to the method developed by Grimmer with the Becke-Johnson damping scheme (D3BJ).\textsuperscript{[9]} The spin-orbit coupling (SOC) operator was treated by a mean-field approximation to the Breit–Pauli operator (SOCType 3 in ORCA).\textsuperscript{[10]} Construction of the effective potential included one-electron terms, the Coulomb term was computed with the RI approximation, exchange was incorporated via one-center exact integrals including spin-other orbit interactions, and local DFT correlation was included with VWN5 (all these choices are set with the keyword "SOCFlags 1,3,3,1” in ORCA).

\subsection*{Computational results}

\begin{figure}[htbp]
	\centering
	\includegraphics[width=0.7\textwidth]{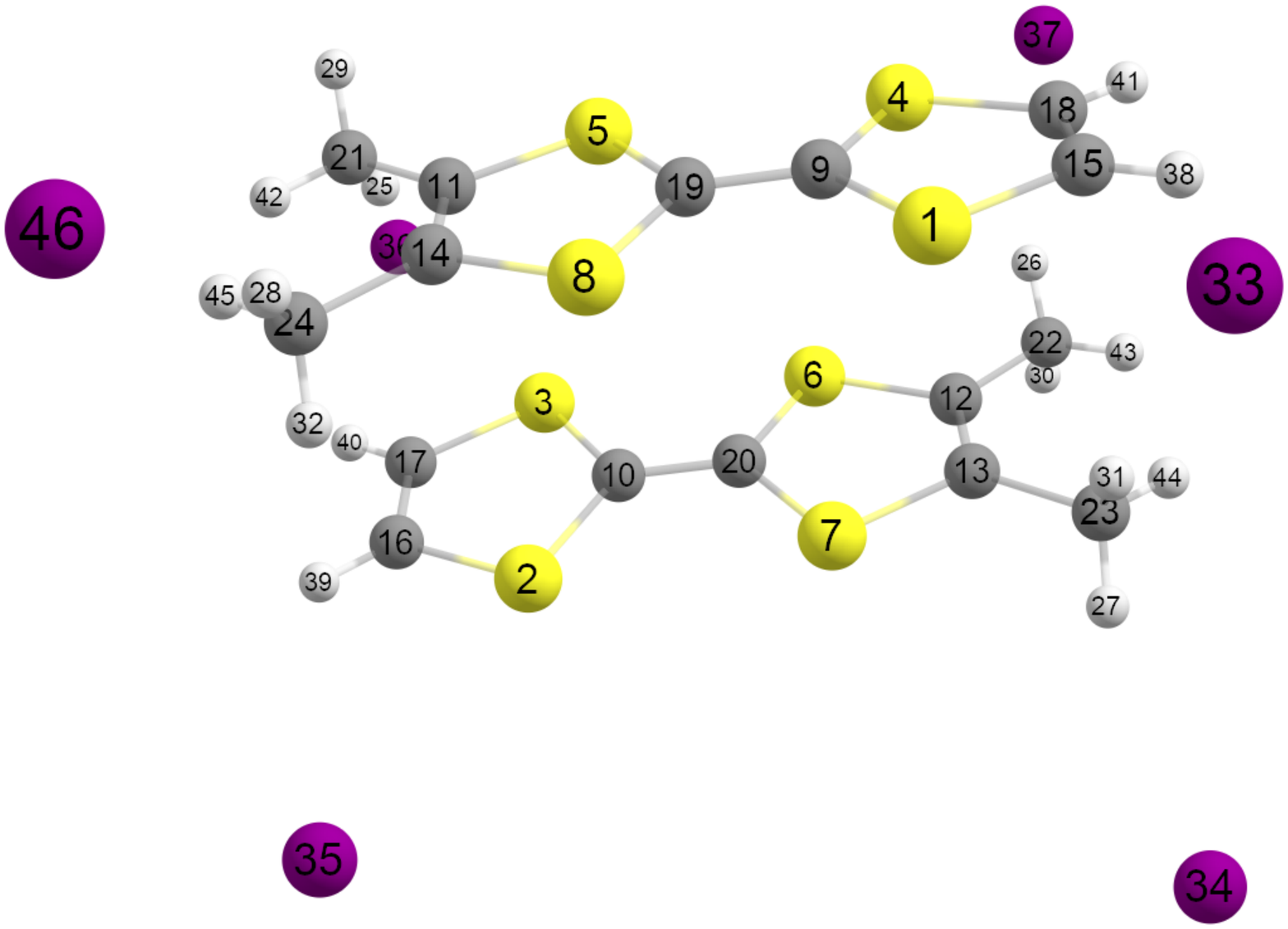} 
	\caption{Atom labelling of the DFT model for \I}
	\label{fig:s13}
\end{figure}

\begin{figure}[htbp]
	\centering

	\includegraphics[width=0.7\textwidth]{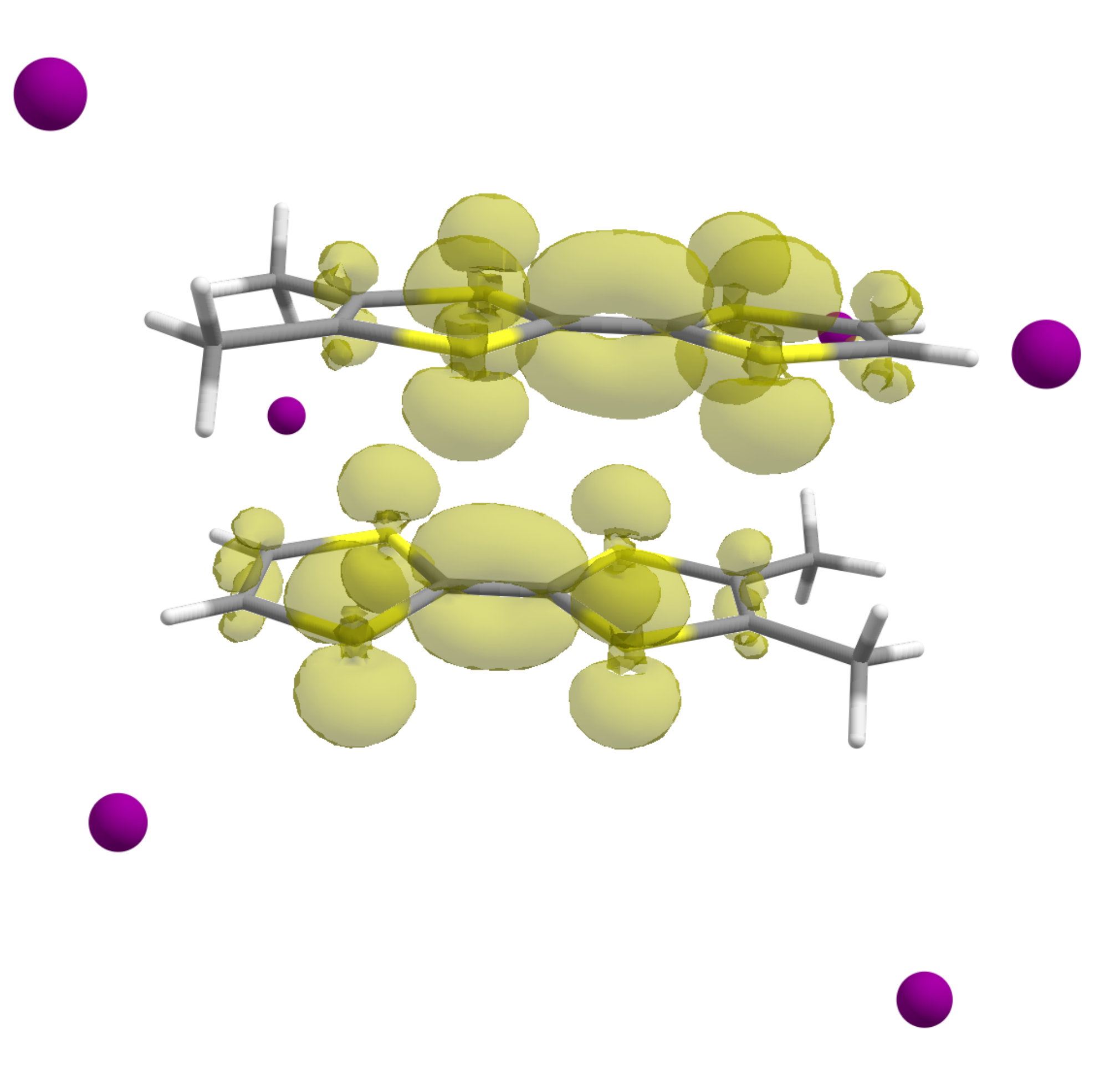} 
	\caption{DFT-computed spin density plot of \I}
	\label{fig:s14}
\end{figure}

\begin{table}[htbp]
	\centering
	\caption{DFT-computed hyperfine coupling constants (total values A$_{ii}$, and individual spin-dipolar components A$_{ii}^{\text{SD}}$, in MHz for \I}
	\label{tab:s1}
	\begin{tabular}{@{}lrrrrrrr@{}}
		\toprule
		Center & A$_{11}$ & A$_{11}^{\text{SD}}$ & A$_{22}$ & A$_{22}^{\text{SD}}$ & A$_{33}$ & A$_{33}^{\text{SD}}$ & A$_{\text{iso}}$ \\
		\midrule
		H25 & 1.23  & -0.74 & 1.74  & -0.23 & 2.95  & 0.97  & 1.97  \\
		H29 & 1.04  & -0.55 & 1.16  & -0.42 & 2.58  & 0.98  & 1.59  \\
		H42 & -0.38 & -0.30 & -0.73 & -0.64 & 0.87  & 0.95  & -0.08 \\
		H28 & 0.09  & -0.52 & 0.25  & -0.37 & 1.53  & 0.89  & 0.62  \\
		H32 & 0.54  & -0.71 & 1.09  & -0.17 & 2.15  & 0.88  & 1.26  \\
		H45 & -0.25 & -0.30 & -0.50 & -0.55 & 0.90  & 0.85  & 0.05  \\
		H38 & 0.44  & 1.66  & -1.83 & -0.53 & -2.41 & -1.13 & -1.26 \\
		H41 & 0.25  & 1.83  & -1.99 & -0.37 & -3.09 & -1.46 & -1.61 \\
		H39 & 0.44  & 1.66  & -1.83 & -0.53 & -2.41 & -1.13 & -1.26 \\
		H40 & 0.25  & 1.83  & -1.99 & -0.37 & -3.09 & -1.46 & -1.61 \\
		H26 & 1.23  & -0.74 & 1.74  & -0.23 & 2.95  & 0.97  & 1.97  \\
		H30 & 1.04  & -0.55 & 1.16  & -0.42 & 2.58  & 0.98  & 1.59  \\
		H43 & -0.38 & -0.30 & -0.72 & -0.64 & 0.87  & 0.95  & -0.08 \\
		H27 & 0.09  & -0.52 & 0.25  & -0.37 & 1.53  & 0.89  & 0.62  \\
		H31 & 0.54  & -0.71 & 1.09  & -0.17 & 2.15  & 0.88  & 1.26  \\
		H44 & -0.25 & -0.30 & -0.50 & -0.55 & 0.90  & 0.85  & 0.05  \\
		\bottomrule
	\end{tabular}
\end{table}

\subsection*{References}

[1] F. Neese \textit{Wiley Interdiscip. Rev.: Comput. Mol. Sci.} \textbf{8}, 1 (2018).

[2] F. Neese, \textit{WIREs Comput. Mol. Sci.} \textbf{12}, e1606, (2022).

[3] A.D. Becke, \textit{J. Chem. Phys.} \textbf{98}, 1372–1377 (1993).

[4] C. Lee, W. Yang, R.G. Parr, \textit{Phys. Rev. B.} \textbf{37}, 785–789 (1988).

[5] P. Pollak, F. Weigend, \textit{J. Chem. Theory Comput.} \textbf{13}, 3696-3705 (2017).

[6] D. Peng, N. Middendorf, F. Weigend, M. Reiher, \textit{J. Chem. Phys.} \textbf{138}, 184105 (2013).

[7] Y. J. Franzke, N. Middendorf, F. Weigend, \textit{J. Chem. Phys.} \textbf{148}, 104110 (2018).

[8] Y. J. Franzke, J. M. Yu, \textit{J. Chem. Theory Comput.} \textbf{18}, 323-343 (2022).

[9] S. Grimme, J. Antony, S. Ehrlich, H. A. Krieg \textit{J. Chem. Phys.} \textbf{132}, 154104 (2010).

[10] S. Grimme, S. Ehrlich, L. Goerigk, \textit{J. Comput. Chem.} \textbf{32}, 1456–1465 (2011).

[11] F. Neese, \textit{J. Chem. Phys.} \textbf{122}, 034107 (2005).

\end{document}